\documentclass{elsart}
\usepackage{graphics,natbib,graphicx,psfig,amssymb,ccaption}
\journal{New Astronomy}
\begin{document}
\begin{frontmatter}
\title{Galactic model parameters of cataclysmic 
variables: Results from a new absolute 
magnitude calibration with 2MASS and $WISE$}
\author[Fen]{A. \"Ozd\"onmez\corauthref{cor}},
\corauth[cor]{Corresponding author. Fax: +90 212 440 03 70}
\ead{aykutozdonmez@gmail.com}
\author[istanbul]{T. Ak},
\author[istanbul]{S. Bilir}
\address[Fen]{Istanbul University, Graduate School of Science and 
Engineering, Department of Astronomy and Space Sciences, Istanbul, Turkey}
\address[istanbul]{Istanbul University, Faculty of Science, Department 
of Astronomy and Space Sciences, 34119 University, Istanbul, Turkey}

\begin{abstract}

In order to determine the spatial distribution, Galactic model 
parameters and luminosity function of cataclysmic variables 
(CVs), a $J$-band magnitude limited sample of 263 CVs 
has been established using a newly constructed 
period-luminosity-colours (PLCs) relation which includes 
$J$, $K_{s}$ and $W1$-band magnitudes in 2MASS and $WISE$ 
photometries, and the orbital periods of the systems.
This CV sample is assumed to be homogeneous regarding to distances 
as the new PLCs relation is calibrated with new or re-measured 
trigonometric parallaxes. Our analysis shows that the scaleheight of 
CVs is increasing towards shorter periods, although selection effects 
for the periods shorter than 2.25 h dramatically decrease the 
scaleheight: the scaleheight of the systems increases from 192 pc 
to 326 pc as the orbital period decreases from 12 to 2.25 h. 
The $z$-distribution of all CVs in the sample is well fitted by an 
exponential function with a scaleheight of 213$^{+11}_{-10}$ pc. 
However, we suggest that the scaleheight of CVs in the Solar 
vicinity should be $\sim$300 pc and that the scaleheights derived 
using the sech$^2$ function should be also considered in the 
population synthesis models. The space density of CVs in the Solar vicinity 
is found 5.58(1.35)$\times 10^{-6}$ pc$^{-3}$ which is in 
the range of previously derived space densities and not 
in agreement with the predictions of the population models. 
The analysis based on the comparisons of the luminosity function of 
white dwarfs with the luminosity function of CVs in this study 
show that the best fits are obtained by dividing the luminosity 
functions of white dwarfs by a factor of 350-450. 
 
\end{abstract}

\begin{keyword}
97.80.Gm Stars: cataclysmic variables  
\sep 97.10.Xq Stars: luminosity function
\sep 98.35.Pr Galaxy: solar neighbourhood   
\end{keyword}

\end{frontmatter}

\section{Introduction}

Cataclysmic variables are defined as short-period semi-detached 
binary stars in which a white dwarf, the primary star, accretes 
matter via a gas stream and an accretion disc from a donor star, 
the secondary. Secondary stars are usually low-mass 
near-main-sequence stars with rare exceptions where the 
secondary evolved via nuclear processes. A bright spot is formed 
by shock-heating where the matter stream impacts the accretion disc. 
An accretion disc can not be formed in magnetic CVs with highly 
magnetized white dwarfs in which matter transfer occurs through 
accretion channels and columns. For comprehensive reviews of CVs, 
see \cite{Wa95}, \cite{Hel01}, \cite{Sm07} and \cite{Kn11}.

As the orbital period ($P_{orb}$) is the most precisely determined 
observational parameter for CVs, properties of the orbital period 
distribution can be easily examined. The orbital period gap between 
roughly 2 and 3 h \citep{SR83,K88,Kn11} and a sharp cut-off at about 
80 min, period minimum \citep{Hametal1988,Wiletal05,Ganetal09}, are 
the most striking features of the CV period distribution. Since this 
distribution is a useful indicator of dynamical evolution of the 
systems, most studies on the evolution of CVs are mainly concentrated 
on the explanation of the orbital period distribution. The standard 
CV formation and evolution theory successfully explains the orbital 
period gap and the period minimum, since its predictions on the 
orbital period gap and the period minimum are supported by 
observations \citep{Patetal05,Ganetal09,Kn11,R12}. It should be 
stated that the evolution of magnetic CVs may be different from 
non-magnetic CVs \citep{To09}. 

In addition to the orbital period distribution, spatial distributions 
and kinematical analysis could be important to test evolutionary 
scenarios and their predictions, as the data obtained from stellar 
statistics must be in agreement with the proposed evolutionary 
schemes \citep{Du84}. Although it is hard to measure the radial 
velocity variation of a CV and such measurements are affected 
by the other components and activity level of the system, stellar 
statistics could give more reliable results depending on the 
completeness of the samples \citep{Aketal08,Aketal10}. However, 
observational selection effects are usually strong, and even a rough 
estimate of Galactic model parameters such as space density and 
scaleheight obtained from the stellar statistics may be crucial 
to constrain the evolutionary models \citep{Pat84}. For example, 
observations of CVs and predictions from CV population models give 
space densities $10^{-7}-10^{-4}$ pc$^{-3}$ 
\citep{Wa74,Pat84,Pat98,Ri93,Schetal02a,ABetal05,Pretal07a,
Aketal08,Revetal08,PrKn12,Pretal13} and $10^{-5}-10^{-4}$ pc$^{-3}$ 
\citep{RB86,dK92,Kol93,Pol96,Wiletal05,Wiletal07}, respectively. 
Although there is a rough agreement between the observed and 
predicted space densities, the wide range of the observational values 
is striking. Similarly, the observed luminosity function \citep{Ri93,
Sazetal06,Revetal08,Aketal08,Bycetal10,PrKn12} of CVs can give 
clues on evolution and mass distribution as a function of the orbital 
period. Regarding the exponential scaleheight of CVs, \cite{Pat84} 
and \cite{vParetal96} found 190$\pm$30 pc and 160-230 pc, 
respectively, while \cite{Pretal07b} adopted 120, 260 and 450 pc 
for long, for normal short orbital period systems and for period 
bouncers, respectively, for modeling the 
Galactic population of CVs. \cite{Ganetal09} concluded that 
scaleheight of CVs is very likely larger than the 190 pc found 
by Patterson (1984). It should be noted that previous measurements 
for the scaleheight of CVs were made from the samples which are 
strongly biased towards bright objects. That is why \cite{Pretal07b} 
argued that the 190 pc used by \cite{Pat84} can be suitable only for 
youngest CVs. However, \cite{Aketal08} found from a large sample of 
CVs including both short and long orbital period systems that the 
exponential scaleheight is 158$\pm$14 pc for the 2MASS 
\citep[Two Micron All Sky Survey;][]{Skrutetal06} $J$-band limiting 
apparent magnitude of 15.8 which was set by authors to obtain 
a complete sample.  

Disagreements of the Galactic model parameters in various studies 
of CVs could originate from selection effects and inadequate numbers 
of systems with reliable distances, which are encountered in almost 
all cases \citep{Aketal08}. With the exception of \cite{Aketal08}, 
previous Galactic model parameters of CVs were derived from the 
samples for which distances were estimated using various methods. 
As the first step of deriving Galactic model parameters and 
a luminosity function is to collect truthful distances of systems in 
a selected sample, it is crucial to find a single method for 
estimating CV distances. Near-infrared photometric methods proposed 
by \cite{Aketal07}, \cite{Beu06} and \cite{Kni06,Kni07} could allow 
researchers to collect a fairly homogeneous sample of CVs regarding 
the distances. All three methods are based on the Barnes-Evans 
relation \citep{Bar76} which states that the surface brightness of 
the late-type main-sequence stars, such as the secondary stars in CVs, 
in near-infrared is nearly constant. An important advantage of 
these methods is the use of magnitudes measured in the near-infrared 
wavelengths for which the interstellar absorption is much weaker than 
that in the optical wavelengths. Several studies to estimate CV 
distances using the Barnes-Evans relation have been done 
\citep{Bai81,Berreta85,Sprtsetal96}. \cite{GarRin12} express that the 
method proposed by \cite{Aketal07} works best overall for all CVs, an 
expected result as the PLCs relation in \cite{Aketal07} is calibrated 
by the trigonometric parallaxes which is the most reliable distance 
estimation method. A new relation similar to Ak et al.'s 
(\citeyear{Aketal07}) PLCs relation may be re-constructed 
using the near-infrared 2MASS and mid-infrared $WISE$ \citep[Wide-field 
Infrared Survey Explorer;][]{Wri10} photometries together that comprise 
a wide range of wavelengths. Such a new relation must be calibrated 
using corrected $Hipparcos$ parallaxes \citep{vLeeu07} and new or 
re-measured trigonometric parallaxes of CVs 
\citep{Pat08,Thoretal08,Thoretal09}. It should be noted that absolute 
magnitudes of the secondary stars in CVs can not be directly calculated 
from the PLCs relation. Although almost all flux in near and 
mid-infrared comes from the secondary star in a CV \citep{Haretal13}, 
it is clear that outer parts of the accretion disc or circumbinary dust 
can contribute to the total infrared flux of the system.

The aim of this paper is to derive the Galactic model parameters and 
luminosity function of CVs in the Solar neighbourhood. As the distance 
is the key parameter for such a study, we first obtain in Section 2 
a new PLCs relation using $J$, $K_{s}$ and $W1$-band magnitudes in 
2MASS and $WISE$ photometries and $P_{orb}$ of CVs with new or 
re-measured trigonometric parallaxes in order to collect a homogeneous 
sample of CVs regarding to distances. Section 3 includes the estimation 
of the Galactic model parameters of CVs in a sample collected from 
Ritter $\&$ Kolb's (\citeyear[][Edition 7.20]{RK03}) catalogue. 
Also, a discussion of the completeness of the sample and the analysis 
regarding to space distribution and luminosity function are given in 
Section 3. We compare and discuss the results of this study in Section 4.

\section{The PLCs relation}

\subsection{The data}

The data sample used in construction of the new PLCs relation consists 
of CVs in Table 1 including their classes, orbital periods ($P_{orb}$), 
trigonometric parallaxes ($\pi$), relative parallax errors 
($\sigma_{\pi}/\pi$), 2MASS $J$ and $K_{s}$ magnitudes and $WISE$ $W1$ 
magnitudes. Their classes and orbital periods were taken from Ritter 
$\&$ Kolb's\footnote[1]{http://physics.open.ac.uk/RKcat/} 
(\citeyear[][Edition 7.20]{RK03}) catalogue. The near-infrared $J$ and 
$K_{s}$ magnitudes of CVs in this calibration sample were taken from 
the point-source catalogue and atlas \citep{Cutetal03} which is based 
on the 2MASS observations. Mid-infrared $WISE$ $W1$-band magnitudes of 
CVs were taken from NASA/IPAC Infrared Science 
Archive\footnote[2]{http://irsa.ipac.caltech.edu/} \citep{Cutetal12}. 
In order to eliminate misidentification, coordinates of the systems 
taken from \cite{RK03} and $WISE$ database were matched with 2MASS 
images in Aladin Sky 
Atlas\footnote[3]{http://aladin.u-strasbg.fr/aladin.gml}. $WISE$ is an 
infrared space telescope with much higher sensitivity than previous 
survey missions (Wright et al., 2010). The $WISE$ space telescope 
surveyed the entire sky from 2010, January 14 to 2010, July 17 in four 
mid-infrared bands (3.4, 4.6, 12 and 22 $\mu$m). These bands are 
denoted as $W$1, $W$2, $W$3 and $W$4 with the angular resolutions 
6.1, 6.4, 6.5 and 12 arcsec, respectively. $WISE$ goes one magnitude 
deeper than the 2MASS $K_{s}$-band magnitude in $W$1 for sources with 
spectra close to that of an A0 star and even deeper for moderately 
red sources like K-type stars \citep{Goketal13}. Note that secondary 
stars in CVs are near, but not necessarily identical to, G-K-M type 
main-sequence stars \citep{Beu98,Kni06}. 

Systems listed in Table 1 have orbital periods 82.4 $\leq P_{orb}$(min) 
$\leq$ 720 as CVs with $P_{orb}>$ 720 and $P_{orb}<$ 82 min have 
secondaries on the way to becoming a red giant \citep{Hel01} and 
a degenerate star \citep{Ganetal09}, respectively.


\begin{table*}
\setlength{\tabcolsep}{2pt}
\scriptsize{
\begin{center}
\caption{CVs with trigonometric parallax measurements. Types and orbital 
periods ($P_{orb}$) were taken from Ritter $\&$ Kolb 
(\citeyear[][Edition 7.20]{RK03}). DN, NL and N denote dwarf nova, 
nova-like star and nova, respectively. $J$ and $K_{s}$ magnitudes were 
collected from the 2MASS point-source catalogue and atlas \citep{Cutetal03}, 
the $WISE$ $W1$ magnitudes from NASA/IPAC Infrared Science Archive 
\citep{Cutetal12}. $\pi$ denotes trigonometric parallax, $\sigma_{\pi}/\pi$ 
relative parallax error, $E(B-V)$ colour excess and $M_{J}$ absolute 
magnitude in $J$-band.}
\begin{tabular}{clccccccccc}
\hline
ID & GCVS-name &Type& $P_{orb}$ &    $\pi$    & $\sigma_{\pi}/\pi$ &        $J$       &    $J-K_{s}$    &   $K_{s}-W1$    & $E(B-V)$ & $M_{J}$  \\
   &           &    &   (hr)    &    (mas)    &                    &                  &                 &                 &          &          \\
\hline
1 & QZ Vir     & DN &   1.412   & 10.20$^{a}$ &      0.12          & 14.771$\pm$0.041 & 0.945$\pm$0.066 &  0.461$\pm$0.058 & 0.011   &  9.80    \\
2 & DW Cnc     & NL &   1.435   &  4.80$^{b}$ &      0.21          & 14.654$\pm$0.031 & 0.621$\pm$0.067 &  0.258$\pm$0.066 & 0.016   &  8.50    \\
3 & VY Aqr     & DN &   1.514   & 11.20$^{a}$ &      0.12          & 15.278$\pm$0.051 & 0.690$\pm$0.103 &  0.515$\pm$0.095 & 0.049   & 10.48    \\
4 & BZ UMa     & DN &   1.632   &  4.90$^{b}$ &      0.22          & 14.824$\pm$0.042 & 0.819$\pm$0.069 & -0.237$\pm$0.063 & 0.031   &  8.24    \\
5 & EX Hya     & NL &   1.638   & 15.50$^{c}$ &      0.02          & 12.274$\pm$0.021 & 0.587$\pm$0.030 &  0.240$\pm$0.031 & 0.022   &  8.21    \\
6 & IR Gem     & DN &   1.642   &  3.00$^{b}$ &      0.37          & 15.218$\pm$0.039 & 0.686$\pm$0.081 &  0.649$\pm$0.076 & 0.049   &  7.56    \\
7 & VV Pup     & NL &   1.674   &  9.30$^{b}$ &      0.13          & 15.553$\pm$0.065 & 1.009$\pm$0.113 &  0.716$\pm$0.097 & 0.014   & 10.39    \\
8 & HT Cas     & DN &   1.768   &  9.00$^{b}$ &      0.12          & 14.703$\pm$0.029 & 0.860$\pm$0.061 &  0.336$\pm$0.060 & 0.034   &  9.44    \\
9 & V893 Sco   & DN &   1.823   &  7.40$^{a}$ &      0.32          & 13.222$\pm$0.023 & 0.540$\pm$0.037 &  0.140$\pm$0.039 & 0.040   &  7.53    \\
10 & SU UMa    & DN &   1.832   &  7.40$^{a}$ &      0.23          & 11.777$\pm$0.018 & 0.107$\pm$0.026 & -1.181$\pm$0.031 & 0.021   &  6.10    \\
11 & MR Ser    & NL &   1.891   &  9.20$^{b}$ &      0.11          & 14.082$\pm$0.024 & 0.728$\pm$0.048 & -0.354$\pm$0.048 & 0.019   &  8.88    \\
12 & AR UMa    & NL &   1.932   & 12.20$^{b}$ &      0.10          & 14.148$\pm$0.022 & 0.882$\pm$0.041 &  0.148$\pm$0.042 & 0.007   &  9.57    \\
13 & YZ Cnc    & DN &   2.083   &  3.34$^{d}$ &      0.13          & 13.166$\pm$0.017 & 0.337$\pm$0.027 &  0.024$\pm$0.032 & 0.021   &  5.76    \\
14 & AM Her    & NL &   3.094   & 13.00$^{a}$ &      0.08          & 11.703$\pm$0.020 & 0.701$\pm$0.028 &  0.434$\pm$0.030 & 0.010   &  7.26    \\
15 & TT Ari    & NL &   3.301   &  0.20$^{e}$ &     39.00          & 10.998$\pm$0.018 & 0.120$\pm$0.026 & -2.294$\pm$0.031 & 0.042   &  2.54    \\
16 & V603 Aql  & N  &   3.317   &  4.96$^{f}$ &      0.49          & 11.700$\pm$0.023 & 0.349$\pm$0.039 &  0.556$\pm$0.044 & 0.127   &  5.07    \\
17 & V1223 Sgr & NL &   3.366   &  1.96$^{g}$ &      0.09          & 12.810$\pm$0.021 & 0.170$\pm$0.037 &  0.127$\pm$0.038 & 0.093   &  4.19    \\
18 & LX Ser    & NL &   3.802   &  0.70$^{e}$ &      5.57          & 13.926$\pm$0.026 & 0.280$\pm$0.045 &  0.177$\pm$0.044 & 0.034   &  3.12    \\
19 & KT Per    & DN &   3.904   &  6.90$^{b}$ &      0.17          & 13.311$\pm$0.021 & 0.686$\pm$0.031 &  0.461$\pm$0.033 & 0.068   &  7.45    \\
20 & U Gem     & DN &   4.246   &  9.96$^{d}$ &      0.04          & 11.651$\pm$0.018 & 0.823$\pm$0.025 &  0.304$\pm$0.029 & 0.006   &  6.63    \\
21 & V405 Peg  & DN &   4.264   &  7.20$^{h}$ &      0.15          & 12.666$\pm$0.018 & 0.852$\pm$0.025 &  0.043$\pm$0.029 & 0.094   &  6.87    \\
22 & SS Aur    & DN &   4.387   &  5.99$^{d}$ &      0.06          & 12.701$\pm$0.017 & 0.700$\pm$0.027 & -0.100$\pm$0.032 & 0.050   &  6.55    \\
23 & MQ Dra    & NL &   4.391   &  6.70$^{b}$ &      0.15          & 14.594$\pm$0.034 & 0.829$\pm$0.063 &  0.375$\pm$0.058 & 0.006   &  8.72    \\
24 & IX Vel    & NL &   4.654   & 10.34$^{f}$ &      0.09          &  9.118$\pm$0.027 & 0.290$\pm$0.032 &  0.076$\pm$0.028 & 0.065   &  4.13    \\
25 & HX Peg    & DN &   4.819   &  3.90$^{e}$ &      1.13          & 13.224$\pm$0.021 & 0.298$\pm$0.039 & -0.268$\pm$0.042 & 0.035   &  6.15    \\
26 & V3885 Sgr & NL &   4.972   &  9.45$^{f}$ &      0.19          &  9.955$\pm$0.023 & 0.339$\pm$0.033 &  0.156$\pm$0.034 & 0.021   &  4.81    \\
27 & RX And    & DN &   5.037   &  0.60$^{e}$ &     20.50          & 12.454$\pm$0.023 & 0.894$\pm$0.034 &  0.537$\pm$0.033 & 0.050   &  1.30    \\
28 & TV Col    & NL &   5.486   &  2.70$^{i}$ &      0.04          & 13.197$\pm$0.024 & 0.505$\pm$0.038 &  0.129$\pm$0.042 & 0.026   &  5.33    \\
29 & RW Tri    & NL &   5.565   &  2.93$^{j}$ &      0.11          & 11.938$\pm$0.019 & 0.478$\pm$0.028 & -0.024$\pm$0.030 & 0.075   &  4.20    \\
30 & RW Sex    & NL &   5.882   &  3.46$^{k}$ &      0.71          & 10.321$\pm$0.023 & 0.251$\pm$0.030 &  0.111$\pm$0.031 & 0.034   &  2.99    \\
31 & AH Her    & DN &   6.195   &  3.00$^{a}$ &      0.50          & 11.806$\pm$0.018 & 0.431$\pm$0.029 &  0.211$\pm$0.032 & 0.032   &  4.16    \\
32 & SS Cyg    & DN &   6.603   &  6.06$^{d}$ &      0.07          &  8.516$\pm$0.009 & 0.217$\pm$0.016 & -0.888$\pm$0.026 & 0.084   &  2.36    \\
33 & Z Cam     & DN &   6.956   &  8.90$^{a}$ &      0.19          & 11.571$\pm$0.025 & 0.715$\pm$0.033 &  0.816$\pm$0.032 & 0.011   &  6.31    \\
34 & RU Peg    & DN &   8.990   &  3.55$^{d}$ &      0.07          & 11.069$\pm$0.018 & 0.605$\pm$0.023 &  0.013$\pm$0.029 & 0.049   &  3.78    \\
35 & AE Aqr    & NL &   9.880   & 11.61$^{f}$ &      0.23          &  9.459$\pm$0.021 & 0.686$\pm$0.030 &  0.155$\pm$0.030 & 0.016   &  4.77    \\
36 & QU Car    & NL &  10.896   &  2.52$^{f}$ &      0.52          & 10.972$\pm$0.018 & 0.265$\pm$0.026 &  0.180$\pm$0.029 & 0.085   &  2.90    \\
\hline
\end{tabular}
\end{center}
{a: \citet{Thors03}, 
b: \citet{Thoretal08}, 
c: \citet{Beuetal03}, 
d: \citet{Harretal04}, 
e: \citet{vanAltetal95}, 
f: \citet{vLeeu07}, 
g: \citet{Beuetal04}, 
h: \citet{Thoretal09}, 
i: \citet{McArtetal01}, 
j: \citet{McArtetal99}
k: \citet{Duer99}.}
}
\end{table*}

\subsection{Colour excesses, intrinsic colours and absolute magnitudes}

Although the CVs in the calibration sample are relatively close to 
the Sun, the total interstellar absorption in the direction of the 
system in question should be taken into account. For the determination 
of the total absorption for $J$, $K_{s}$ and $W1$-bands, the equations 
of \cite{FiMu03} and \cite{Biletal11} were used, respectively, i.e. 
$A_{J}=0.887\times E(B-V)$, $A_{K_s}=0.382\times E(B-V)$ and 
$W1=0.158\times E(B-V)$. Here, $E(B-V)=A_{V}/3.1$ is adopted 
\citep{Cardeletal1989}. Colour excesses $E(B-V)$ were obtained from the 
Schlafly \& Finkbeiner's (\citeyear{Schletal11}) maps, which is based on 
the maps of \cite{Schlegeletal1998}, by using the NASA/IPAC Extragalactic 
Database\footnote[4]{http://ned.ipac.caltech.edu/}. As the colour 
excesses $E(B-V)$ in the directions of the stars are actually given up to 
the edge of the Galaxy by the \cite{Schletal11} maps, reduction of 
$E(B-V)$ values from the maps according to the actual distance of each 
system in the calibration sample and estimation of the total interstellar 
absorption in the photometric band used were done as described in 
\cite{Aketal07}. After computing total interstellar absorptions 
$A_{J}$, $A_{K_s}$ and $A_{W1}$ in the direction of the star, the 
de-reddened colours $(J-K_{s})_{0}$ and $(K_{s}-W1)_{0}$ were obtained 
from $J_{0}=J-A_{J}$, $(K_{s})_0=K_{s}-A_{K_{s}}$ and $W1_{0}=W1-A_{W1}$.

Once de-reddened $J$-band apparent magnitudes and distances $d=1/\pi$ were 
obtained for the CVs in the calibration sample, their $J$-band absolute 
magnitudes were easily calculated from the well-known Pogson's equation, 
i.e. $M_{J}=J_{0}-5\log d +5$. The calculated $M_{J}$ values 
are given in Table 1.

\subsection{The PLCs relation}

The data of the 36 systems in Table 1 were collected to derive an 
absolute magnitude calibration, the PLCs relation, for CVs. By checking 
standard deviations and correlation coefficients of the fit equations, after 
trying various colours and equation forms, we preferred to use a fit 
equation in the following form to find the dependence of the absolute 
magnitude $M_{J}$ on the orbital period $P_{orb}$ and colours 
$(J-K_{s})_{0}$ and $(K_{s}-W1)_{0}$: 

\begin{equation}
M_{J}=a+b~\log P_{orb}(h)+c~(J-K_{s})_{0}+d~(K_{s}-W1)_{0}.
\end{equation}

However, as can be seen from Table 1, the relative parallax errors 
$\sigma_{\pi}/\pi$ of some CVs are too high and these systems must not be 
taken into account in the least square fit procedure. Thus, we preferred to 
exclude CVs with relative parallax errors $\sigma_{\pi}/\pi >$ 0.5 
from the calibration sample in Table 1. These systems are TT Ari, 
LX Ser, HX Peg, RX And, RW Sex and QU Car. In addition, we found using 
the data in the AAVSO\footnote[5]{http://www.aavso.org/} (American Association 
of Variable Star Observers) archive that the 2MASS colour indices of SS Cyg 
and SU UMa shift to smaller values during outburst and superoutburst 
activities, respectively. For this reason, we also removed SS Cyg and 
SU UMa from the calibration sample. During the regression analysis, MQ Dra, 
IR Gem and VY Aqr were also removed from the calibration sample due to 
high scatter. It is known that low accretion rate polars like MQ Dra have 
lower luminosities compared to CVs with a similar orbital period 
\citep{Schetal02b,Schmetal05}. The mass of the white dwarf in IR Gem is 
somewhat higher than those in CVs whose orbital periods are near this 
system's orbital period. Thus, the irradiation of the secondary star by the 
white dwarf in IR Gem could be higher than in other CVs near its orbital 
period \citep{Fuetal04}. Regarding VY Aqr, various studies assigned 
different spectral types to the secondary star in this system 
\citep{Haretal09}. Consequently, 25 systems remained in the calibration 
sample. A least square fit gives the coefficients and their 
$\pm$1$\sigma$ errors listed in Table 2. The correlation coefficient 
and standard deviation of the calibration were calculated as $R=0.980$ 
and $\sigma=\pm0.431$ mag, respectively. This relation is valid for the 
ranges; $1.37 \leq P_{orb}$(h) $\leq 12$, $0.13 \leq (J-K_{s})_{0} \leq 1.01$, 
$-0.36 \leq (K_{s}-W1)_{0} \leq 0.82$ and $2.9 < M_{J} < 10.4$ mag. 
Monte Carlo simulations performed to estimate the errors of the 
coefficients suggest that the relation provides absolute magnitudes within 
an accuracy of about $\pm0.29$ mag.


\begin{table}
\begin{center}
\caption{Coefficients and their errors for the PLCs relation derived 
in this study.}
\begin{tabular}{ccccc}
\hline
Coefficient &     $a$      &       $b$     &      $c$     &       $d$      \\
\hline
Value       &   {5.966}    &   {-4.781}    &    {5.037}   &     {0.617}     \\
Error    & {$\pm$0.365} &  {$\pm$0.359} & {$\pm$0.418} &  {$\pm$0.331}   \\
\hline
\end{tabular}
\end{center}
\end{table}

\begin{figure}
\begin{center}
\includegraphics[scale=0.25, angle=0]{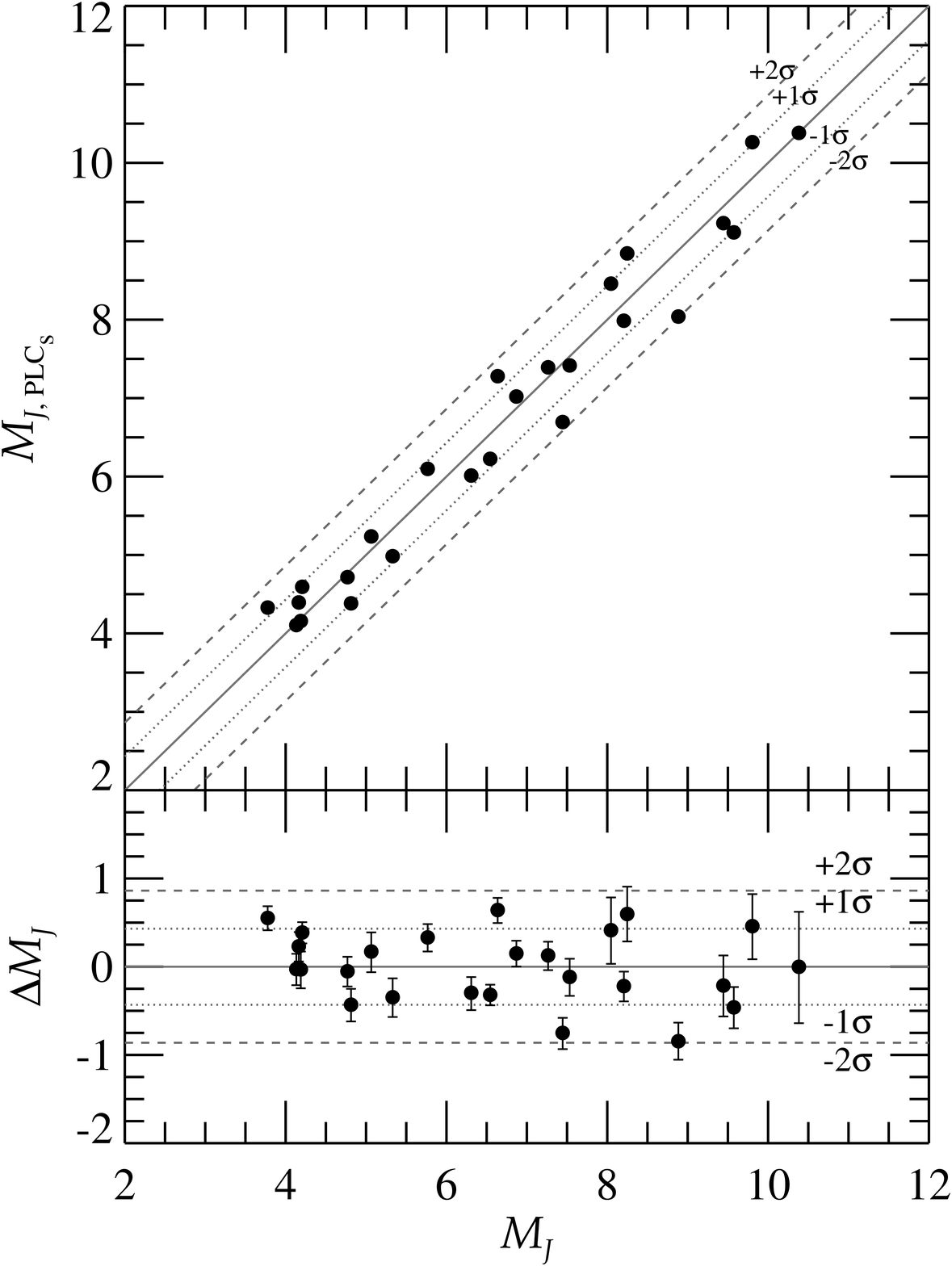}
\caption[] {\small A comparison of the absolute magnitudes calculated from the PLCs relation ($M_{J,PLCs}$) with 
$M_{J}$ magnitudes estimated using trigonometric parallaxes in Table 1. Dotted and dashed lines show 1$\sigma$ and 
2$\sigma$ limits of the PLCs relation, respectively. The bottom panel displays the difference 
$\Delta M_{J}=M_{J,PLCs}-M_{J}$ with $\pm$1$\sigma$ and $\pm$2$\sigma$ limits of the PLCs relation. The solid diagonal 
line in the upper panel represents the equal values.}
\end{center}
\end{figure}

A comparison of the $J$-band absolute magnitudes ($M_{J}$) estimated 
using trigonometric parallaxes with those obtained from the PLCs relation 
($M_{J,PLCs}$) found above is presented in Fig. 1. This figure shows that 
almost all absolute magnitudes estimated from the PLCs relation are within 
the $\pm$1$\sigma$ limit, and the scatter is not more than $\pm$0.9 mag 
despite large errors of faint objects. In addition, absolute magnitudes 
estimated from both method are in agreement for short period CVs and there 
is no systematic effect. A numerical comparison between the PLCs relations 
found in this study and \cite{Aketal07} can be done using the mean values and 
standard deviations of the differences $\Delta M_{J}=M_{J,PLCs}-M_{J}$, which 
are calculated $<\Delta M_{J}>=$-0.005 and $\sigma_{\Delta M_{J}}=\pm$0.403 
mag for this study and $<\Delta M_{J}>=$0.041 and 
$\sigma_{\Delta M_{J}}=\pm$0.732 mag for the PLCs relation in \cite{Aketal07}. 
This comparison shows that the PLCs relation suggested in this study gives 
more reliable results than that in \cite{Aketal07} as the new PLCs relation 
provides $J$-band absolute magnitudes $\sim$2 times more precise compared 
to those of \cite{Aketal07}. We conclude from these comparisons that our PLCs relation 
can be used to collect a homogeneous sample of CVs regarding to the distances. 
Reliable Galactic model parameters of CVs can be derived from such a sample.

\section{Analysis}

\subsection{The data sample}

In order to obtain a sample of CVs, a preliminary list of the systems whose 
orbital periods are known and between 82.4 and 720 min is taken from 
Ritter $\&$ Kolb (\citeyear[][Edition 7.20]{RK03}). The number of CVs in this 
preliminary sample is 858. However, the number of systems having both 2MASS 
and $WISE$ observations in this sample is 509. In addition, 196 of these 
systems were removed because they are beyond the limits of applicability of 
the PLCs relation found above. Thus, the number of CVs that can be used in 
this study reduced to 313 and all are listed in an electronic table as 
a supplement to this paper. In this table, DN denotes dwarf novae, NL 
nova-like stars, N novae, RN recurrent novae. Polars and intermediate-polars 
are indicated with P and IP, respectively. 
In this study, magnetic systems (polars and intermediate polars) were 
evaluated separately since their evolution may be different from 
non-magnetic CVs \citep{To09}. Interstellar absorptions in $J$, $K_{s}$ 
and $W1$ bands were estimated using an iterative process described in 
\cite{Aketal08}. De-reddened colours $(J-K_{s})_{0}$, $(K_{s}-W1)_{0}$ were 
calculated as described in Section 2. Finally, absolute magnitudes in 
$J$-band $M_{J,PLCs}$ and distances $d$ were estimated from the PLCs 
relation and Pogson's equation, respectively.

\subsection{Completeness of the sample}
Incompleteness of surveys could be often expressed as the reason for 
disagreements of the results obtained from observational and theoretical 
studies. Incompleteness of the samples are mostly due to the observational 
selection effects \citep{DelVal96,Aungetal02,Ganetal05,Pretal07b}.

Low-mass transfer rate systems are likely under-represented in the CV 
samples since systems with rare and low-amplitude outbursts are harder 
to discover \citep{Pat98,Pretal07b,Aketal08}. As the theory predicts that 
the most CVs should be intrinsically faint objects, one of the goals 
of CV surveys has been to reduce the selection effects by observing the 
fainter objects. It seems that apparent magnitude limits of surveys are 
one of the strongest selection effects. Therefore, the completeness limit 
of the data must be taken into account in a study based on stellar 
statistics. In order to set a limiting magnitude for the CV sample in 
this study, the histogram for the de-reddened $J$-band magnitudes ($J_{0}$) 
of CVs in the electronic table is shown in Fig. 2. From Fig. 2, 
the apparent bright and 
faint limiting $J$-band magnitudes of 10 and 16 mag were selected, 
respectively, in order to obtain a complete CV sample in a certain volume 
with the Sun in its centre. By removing systems beyond the limiting 
magnitudes, the number of CVs drops to 263 from which the Galactic model 
parameters will be obtained. In Fig. 3, the absolute magnitudes $M_{J}$ of 
CVs in the electronic table is plotted against to their distances. Systems between the 
bright and faint limiting magnitudes are shown with filled circles in 
Fig. 3, while removed CVs are indicated with open circles.

\begin{figure}
\begin{center}
\includegraphics[scale=0.25, angle=0]{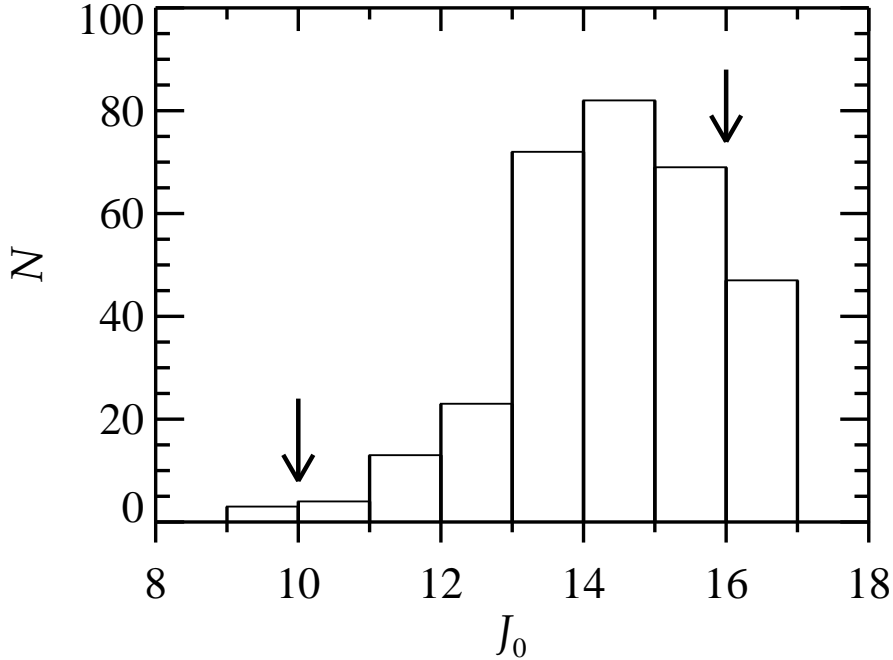}
\caption[] {\small The histogram of the de-reddened $J$-band apparent 
magnitudes ($J_{0}$) of CVs in the electronic table. Bright and faint apparent 
magnitude completeness limits of 10 and 16 mag, respectively, 
were selected using the histogram. Arrows in the histogram indicate 
these limits.}
\end{center}
\end{figure}

\begin{figure}
\begin{center}
\includegraphics[scale=0.25, angle=0]{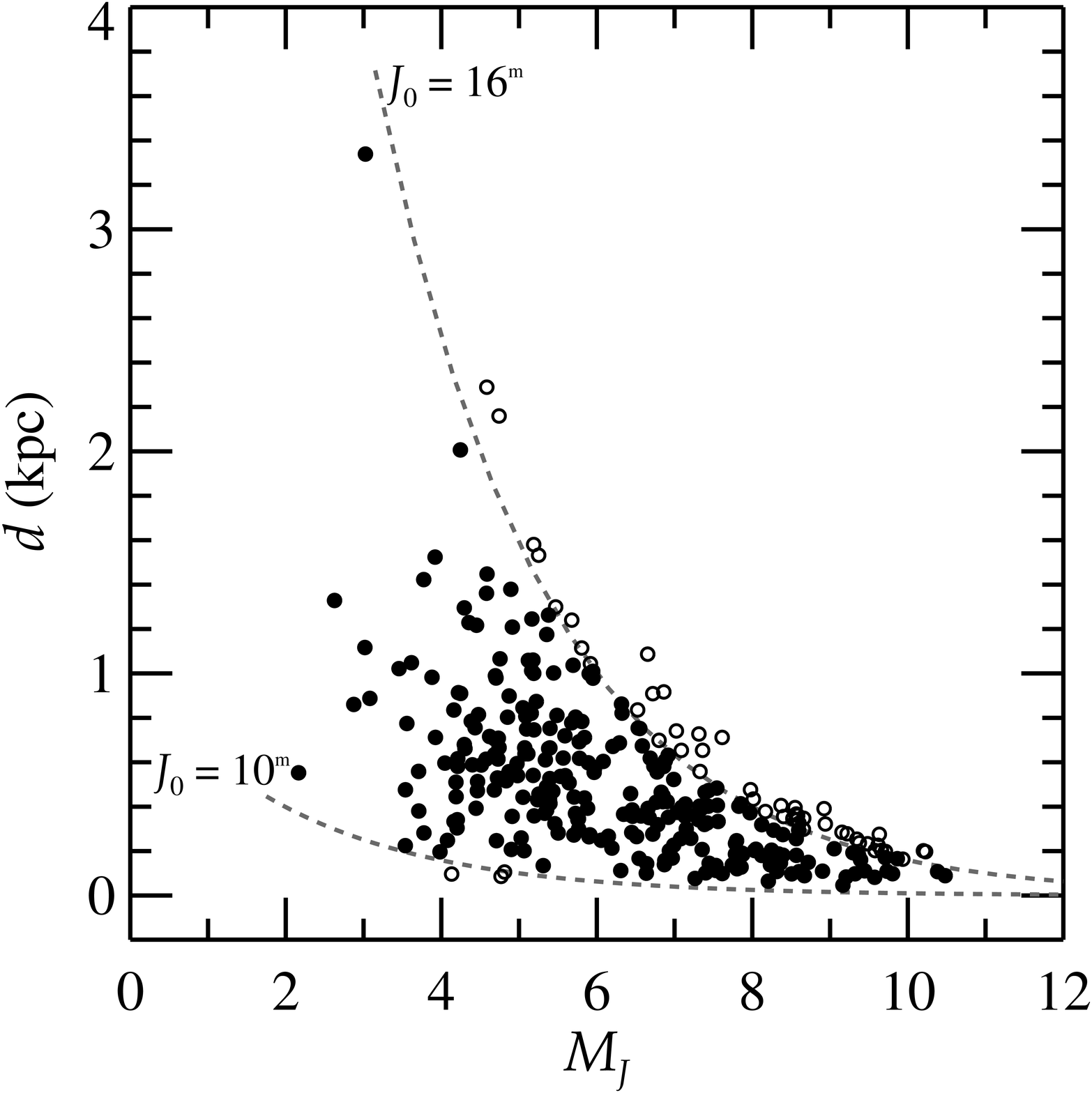}
\caption[] {\small The absolute magnitudes $M_{J}$ of CVs in the electronic table against their 
distances. Dashed lines show bright and faint de-reddened limiting magnitudes in 
$J$-band ($J_0$), while open circles represent the systems beyond the limiting magnitudes.}
\end{center}
\end{figure}

\begin{figure}
\begin{center}
\includegraphics[scale=0.20, angle=0]{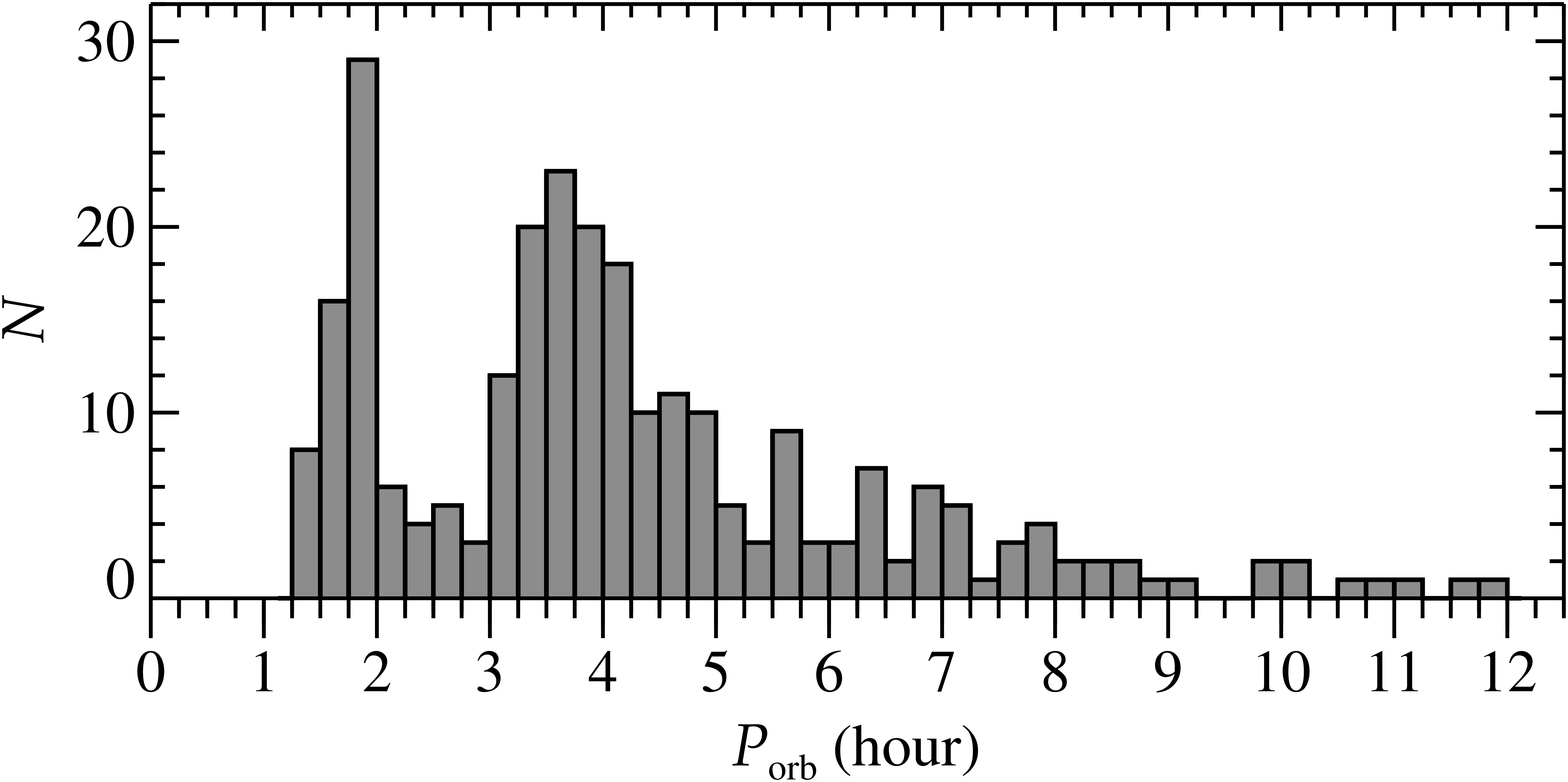}
\caption[] {\small The orbital periods of CVs in the final sample of 263 systems.}
\end{center}
\end{figure}

The orbital period distribution of CVs in the final sample is presented 
in Fig. 4. Among the 263 systems, 124 are nova-like stars, 118 dwarf 
novae, 14 novae, 6 CVs with unknown type and 1 recurrent nova. There are 
73 magnetic CVs (polars and intermediate polars) in this sample. Systems 
that are not classified as polar or intermediate polar are denoted as 
non-magnetic systems. Fig. 4 shows that our sample comprises CVs from 
all orbital period intervals. 

The refined sample of 263 CVs in this study is not free from the 
selection effects. However, selection effects on this sample are likely less 
than for any sample so far in the literature as the limits defined above make 
it almost complete in a certain volume. In addition, it is one of the 
largest CV samples used for deriving the Galactic model parameters available 
in the literature. Thus, Galactic model parameters of CVs in the Solar 
neighbourhood derived from this sample should be reliable and useful for 
CV population models.

\subsection{Spatial distribution}

The distribution of CVs according to equatorial ($\alpha$, $\delta$) and 
Galactic coordinates ($l$, $b$) are plotted in Fig. 5 which indicates that 
the systems in general are symmetrically distributed about the Galactic plane. 
The heliocentric rectangular Galactic coordinates ($X$ towards Galactic 
centre, $Y$ Galactic rotation, $Z$ north Galactic Pole) are very useful in 
order to inspect the Galactic distribution of CVs in the Solar neighbourhood. 
Thus, the Sun centered rectangular Galactic coordinates of CVs in the sample 
were calculated and their projected positions on the Galactic plane ($X, Y$ 
plane) and on a plane perpendicular to it ($X, Z$ plane) are displayed in 
Fig. 6. The numbers, median distances and median heliocentric Galactic 
distances of CVs are given in Table 3. Fig. 6 and Table 3 demonstrate that 
the Galactic positions of CVs in the Solar neighbourhood do not introduce 
a striking bias for this study, in general. However, it should be noted 
that the subgroups such as novae (N), recurrent novae (RN) and CVs with 
unknown type (CV) including only a few members may introduce considerable 
bias to derivation of the Galactic model parameters for these sub-classes. 
The numbers, median distances and median heliocentric Galactic distances 
of CVs above ($P_{orb}$(h) $\geq 3.18$) and below ($P_{orb}$(h) $\leq 2.15$) 
the orbital period gap are also given in Table 3. The lower and upper 
borders for the period gap were adopted from \cite{Kni06}. 

The numbers, median distances and median heliocentric Galactic distances 
of CVs above ($P_{orb}$(h) $\geq 3.18$) and below ($P_{orb}$(h) $\leq 2.15$) 
the period gap are also given in Table 3.

\begin{figure}
\begin{center}
\includegraphics[scale=0.22, angle=0]{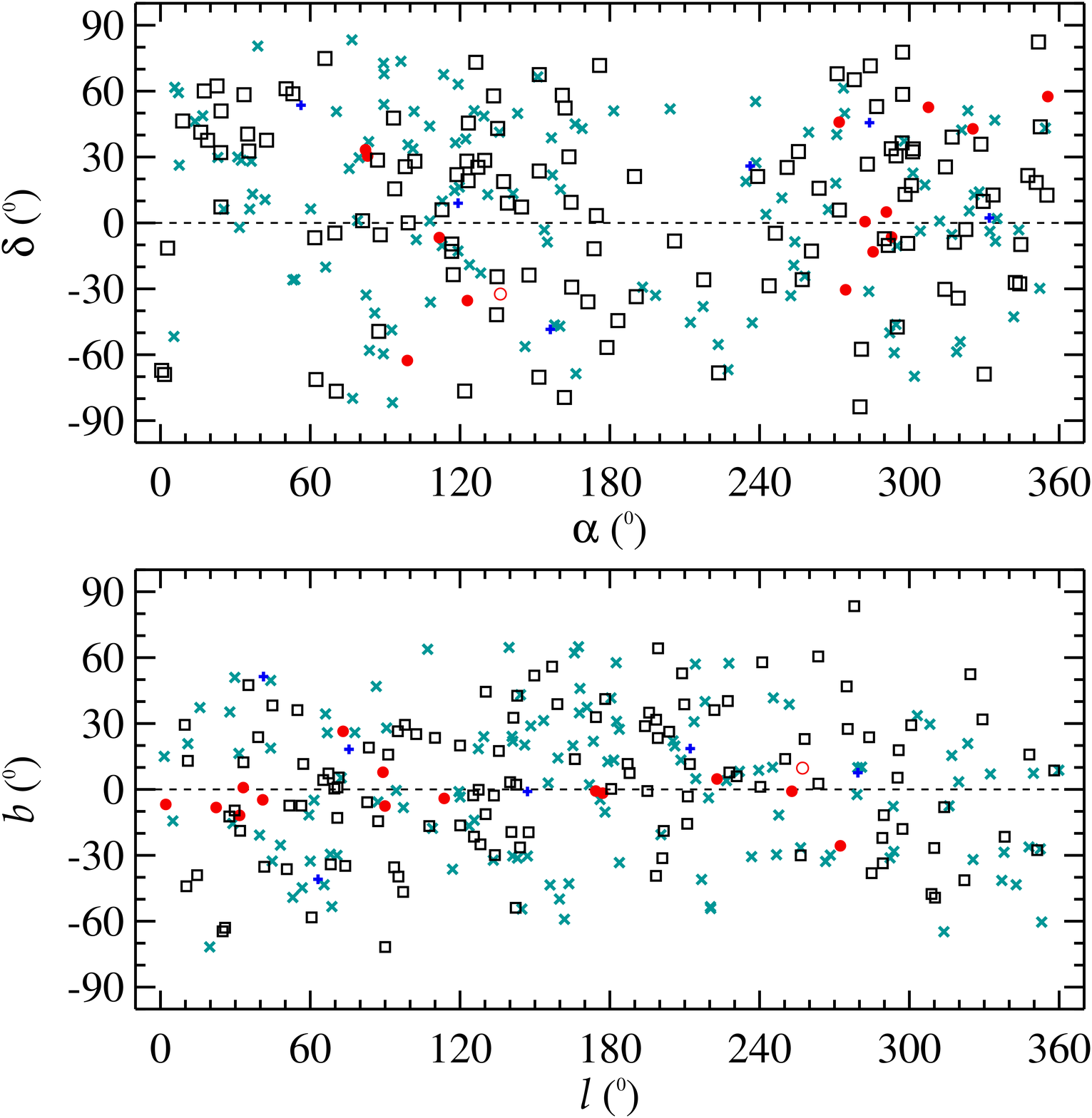}
\caption[] {\small The distribution of CVs according to equatorial 
($\alpha$, $\delta$) and Galactic ($l$, $b$) coordinates. The symbol 
($\square$) denotes dwarf novae, cyan-coloured ($\times$) nova-like stars, 
red-coloured ($\bullet$) novae, red-coloured ($\circ$) recurrent novae 
and blue-coloured ($+$) CVs of unknown type.}
\end{center}
\end{figure}

\begin{figure}
\begin{center}
\includegraphics[scale=0.25, angle=0]{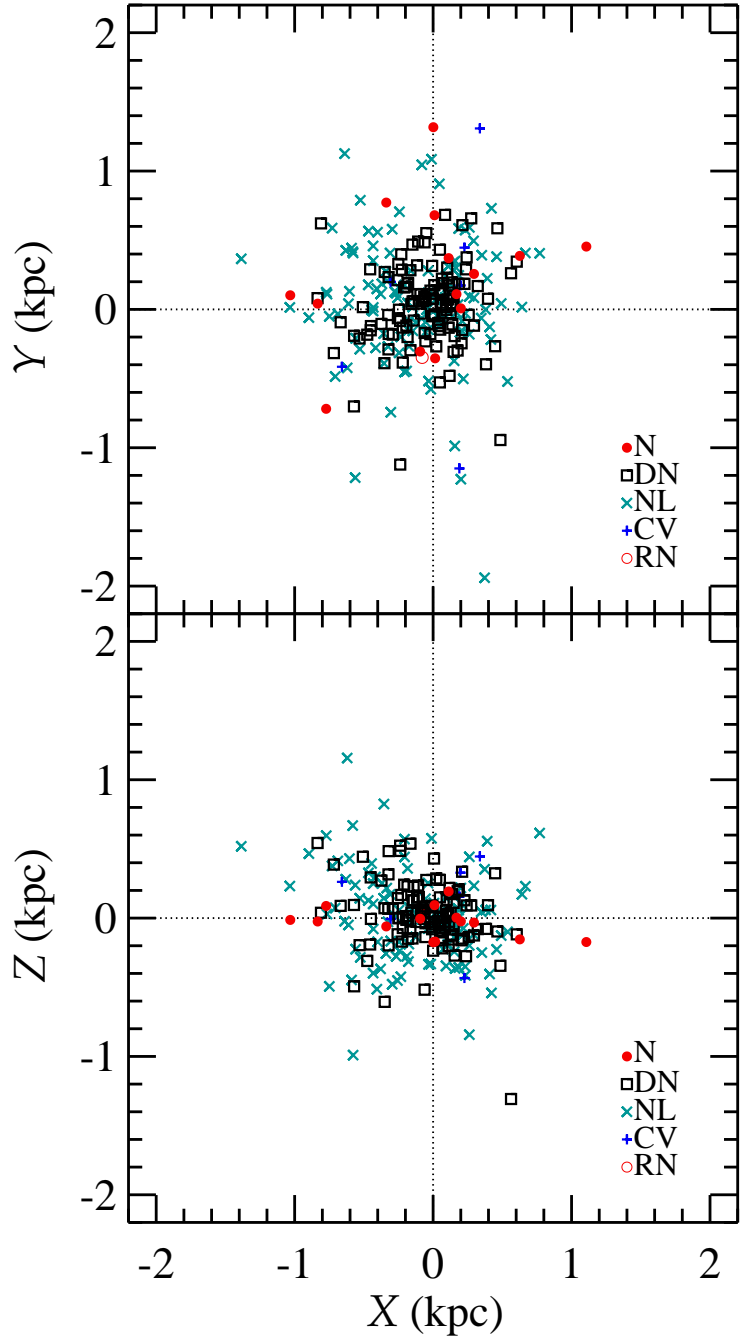}
\caption[] {\small The spatial distribution of CVs in the sample with 
respect to the Sun. $X, Y$ and $Z$ are the Sun centered rectangular 
Galactic coordinates. DN denotes dwarf novae, NL nova-like stars, 
N novae, RN recurrent novae, CV unknown types.}
\end{center}
\end{figure}

\begin{table}
\begin{center}
\caption{The numbers, median distances ($d$) and heliocentric rectangular Galactic 
coordinates ($X,Y,Z$) of all (ALL) CVs in the sample, all CVs below the period gap 
($P_{orb}\leq 2.15$h), all CVs above the period gap ($P_{orb}\geq 3.18$h), dwarf 
novae (DN), nova-like stars (NL), novae (N), magnetic CVs (mCV) and non-magnetic CVs 
(non-mCV). The unclassified 6 CVs and one recurrent nova are not included in subgroups.}
\begin{tabular}{lccccc}
\hline
Type                            & Number & $d$  &   $X$  &   $Y$  &   $Z$  \\
                                &        & (pc) &   (pc) &   (pc) &   (pc) \\
\hline
ALL                             & 263    & 423  &  -36   &   59   &   9   \\
\hline
$P_{orb}\leq 2.15$h             & 59     & 204  &   9    &  -24   &   15  \\
$P_{orb}\geq 3.18$h             & 185    & 540  &  -82   &  102   &    9  \\
\hline
DN                              & 118    & 337  &  -13   &   53   &   12  \\
NL                              & 124    & 528  &  -68   &   45   &   15  \\
N                               & 14     & 720  &   12   &  183   &  -24  \\
\hline
mCV                             & 73     & 385  &  -56   &   17   &   34  \\
non-mCV                         & 190    & 458  &  -20   &   77   &    2  \\
\hline
\end{tabular}
\end{center}
\end{table}

\subsection{Galactic model parameters and space density}

The number of stars per unit volume is a function of the Galactic positions of stars. Studies 
based on the deep sky surveys showed that scalelength of the thin-disc stars is larger than 
2.6 kpc \citep{Biletal06a,Juretal08}. An inspection of the electronic table shows that the distances of 95 per cent of 
CVs in the sample are smaller than 1 kpc. This implies that almost all CVs in the sample should be members 
of the thin-disc component of the Galaxy. Thus, they were not classified according to the 
population types and estimation of scalelength of CVs were not attempted.

$z$-histograms that exhibit the vertical distribution of objects in the Galaxy must 
be studied in order to find the Galactic model parameters of the objects in a sample. 
$z$ is the distance of objects from the Galactic plane: 
$z=z_{\odot}+d\sin\mid b\mid$ where $z_{\odot}$ is the Sun's vertical distance from the 
Galactic plane \citep[24 pc,][]{Juretal08}, $d$ distance from the Sun and $b$ Galactic 
latitude. Hence, $z$-histograms binned per 100 pc for all CVs (ALL), non-magnetic 
systems (non-MCV) and magnetic systems (MCV) in the sample are shown in Fig. 7. 
$z$-histograms of dwarf novae (DN), nova-like stars (NL) and novae (N) are 
presented in Fig. 8, as well. Although the exponential function has usually been used to 
describe the number density variation of stars by the distance from the Galactic plane 
and, thus, to derive the Galactic model parameters, \cite{Biletal06a,Biletal06b} showed that observed 
vertical distribution in the Solar neighbourhood is well-approximated by a secans hyperbolic 
function. Hence, in order to derive scaleheight and number density of CVs in the Solar 
neighbourhood, the following two functions were preferred and fitted to $z$-histograms:

\begin{equation}
n(z)=n_{0}\exp\Biggl(-\frac{\mid z\mid}{H}\Biggr),
\end{equation}

and 

\begin{equation}
n(z)=n_{0}~{\rm sech}^{2}\Biggl(-\frac{\mid z\mid}{H_z}\Biggr),
\end{equation}

where $n_{0}$ is the number of stars in the Solar neighbourhood, $H$ and $H_z$ are the exponential 
and sech$^2$ scaleheights, respectively. By definition, the relation between the exponential and 
sech$^2$ scaleheight is $H=1.08504\times H_z$ \citep{Biletal06a,Biletal06b}. All error estimates in the 
analysis were obtained by changing Galactic model parameters until an increase or decrease 
by $\pm$1$\sigma$ in $\chi^{2}$ was achieved \citep{Pressetal1997}. The best fits to the 
$z$-histograms of all CVs in the sample and subgroups are shown in Figs. 7 and 8. The 
scaleheight $H$ and the number of stars in the Solar neighbourhood $n_{0}$ obtained from the 
minimum $\chi^{2}$ analysis are given in Table 4. In the last column of Table 4, the observed 
number of CVs with $z\leq$ 100 pc ($N_{obs}$) is also listed.

\begin{figure}
\begin{center}
\includegraphics[scale=0.25, angle=0]{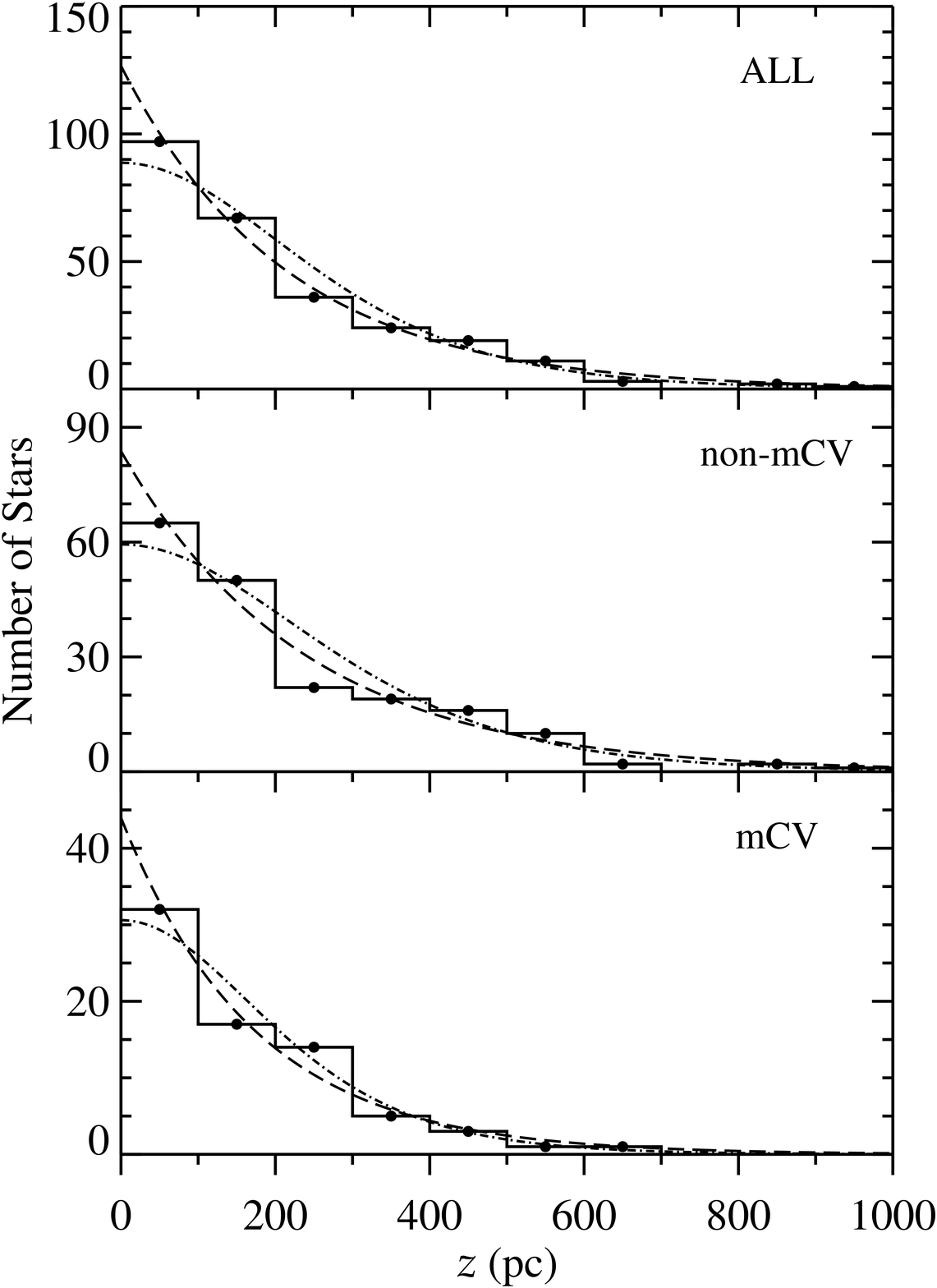}
\caption[] {\small The $z$-histograms for CVs with $10 \leq J_{0} \leq 16$ mag. 
ALL denotes all systems in the sample, non-MCV non-magnetic systems and mCV magnetic 
systems. The dashed line represents the exponential function, the dot-dashed line 
the sech$^2$ function.}
\end{center}
\end{figure}

\begin{figure}
\begin{center}
\includegraphics[scale=0.25, angle=0]{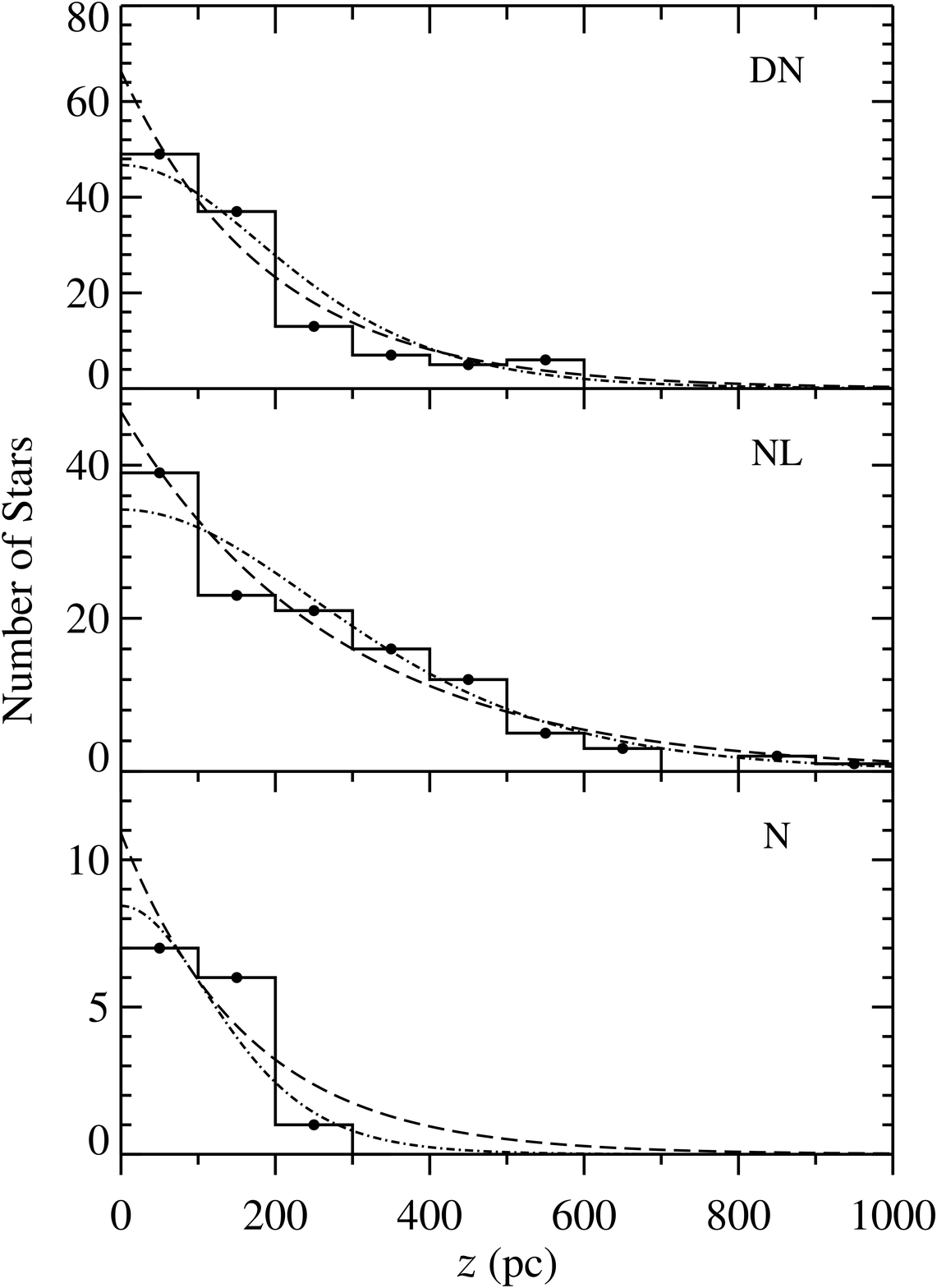}
\caption[] {\small The $z$-histograms for dwarf novae (DN), nova-like stars (NL) and novae (N) in 
the sample with $10 \leq J_{0} \leq 16$ mag. Lines are as described in Fig. 7.}
\end{center}
\end{figure}

\begin{table}
\center
{\scriptsize
\caption{{\scriptsize The Galactic model parameters for CVs in the sample. Functions used in the 
models are given in the second column. Here, $n_{0}$ is the number of stars in the Solar 
neighbourhood, $H$ (pc) the exponential scaleheight for the function given in the second 
column, $\chi^{2}_{min}$ the minimum $\chi^{2}$ value. Denotes for subgroups are as in Table 3. 
$N_{obs}$ is the observed number of systems with $z \leq$ 100 pc.}}
\begin{tabular}{lccccc}
\hline
Subgroup  &  Function        &  $n_{0}$        &     $H$             & $\chi^{2}_{min}$ &  $N_{obs}$  \\
\hline
ALL        &  $\exp$        &  127$^{+8}_{-8}$ &  213$^{+11}_{-10}$  & 3.45             &  97  \\    
           &  sech$^2$      &  89$^{+6}_{-5}$  &  326$^{+13}_{-12}$  & 7.37             &  97  \\    
\hline
DN         &  $\exp$        &  66$^{+6}_{-6}$  &  191$^{+16}_{-14}$  & 5.82             &  49  \\    
           &  sech$^2$      &  47$^{+4}_{-4}$  &  289$^{+18}_{-15}$  & 9.22             &  49  \\    
\hline
NL         &  $\exp$        &  47$^{+4}_{-4}$  &  278$^{+25}_{-22}$  & 3.27             &  39  \\    
           &  sech$^2$      &  34$^{+3}_{-3}$  &  404$^{+26}_{-23}$  & 3.42             &  39  \\    
\hline
N          &  $\exp$        &  11$^{+3}_{-2}$  &  164$^{+79}_{-41}$  & 1.54             &  7  \\    
           &  sech$^2$      &   9$^{+3}_{-2}$  &  196$^{+64}_{-35}$  & 0.97             &  7  \\    
\hline
non-mCV    &  $\exp$        &  84$^{+6}_{-5}$  &  236$^{+6}_{-5}$    & 6.26             &  65  \\    
           &  sech$^2$      &  59$^{+4}_{-4}$  &  356$^{+17}_{-15}$  & 9.03             &  65 \\    
\hline
mCV        &  $\exp$        &  44$^{+5}_{-5}$  &  173$^{+18}_{-15}$  & 1.93             &  32  \\    
           &  sech$^2$      &  31$^{+4}_{-3}$  &  264$^{+21}_{-17}$  & 1.97             &  32 \\  
\hline
\end{tabular}
}
\end{table}

\begin{table}
\center
{\scriptsize
\caption{{\scriptsize The Galactic model parameters for CVs in terms of the orbital period. $n_{0}$, $H$ (pc)
and $\chi^{2}_{min}$ are as defined in Table 4. ALL denotes all CVs in the subgroup, mCV magnetic CVs and non-mCV 
non-magnetic CVs. $N_{obs}$ is the observed number of systems with $z \leq$ 200 pc.}}
\begin{tabular}{lccccc}
\hline
Subgroup  &  Function     &  $n_{0}$      &     $H$          & $\chi^{2}_{min}$ &  $N_{obs}$  \\
\hline
$1.37 \leq P_{orb}$(h) $< 2.25$ (ALL) &  $\exp$       &  97$^{+13}_{-12}$&  135$^{+14}_{-12}$  & 0.39             &  47  \\    
                                    &  sech$^2$       &  57$^{+8}_{-7}$  &  230$^{+20}_{-16}$  & 1.26             &  47 \\  
non-mCV                             &  $\exp$         &  65$^{+11}_{-10}$&  131$^{+17}_{-13}$  & 0.62             &  31  \\    
                                    &  sech$^2$       &  37$^{+7}_{-6}$  &  226$^{+25}_{-19}$  & 1.40             &  31 \\  
mCV                                 &  $\exp$         &  32$^{+8}_{-6}$  &  144$^{+28}_{-20}$  & 0.01             &  16  \\    
                                    &  sech$^2$       &  19$^{+5}_{-4}$  &  239$^{+39}_{-28}$  & 0.10             &  16 \\  
\hline
$2.25 \leq P_{orb}$(h) $< 3.7$ (ALL) &  $\exp$        &  41$^{+6}_{-5}$  &  326$^{+49}_{-38}$  & 2.86             &  30  \\    
                                    &  sech$^2$       &  31$^{+4}_{-4}$  &  444$^{+49}_{-38}$  & 2.58             &  30 \\  
non-mCV                             &  $\exp$         &  27$^{+4}_{-4}$  &  364$^{+73}_{-53}$  & 1.72             &  20  \\    
                                    &  sech$^2$       &  20$^{+3}_{-3}$  &  479$^{+69}_{-50}$  & 1.43             &  20 \\  
mCV                                 &  $\exp$         &  13$^{+4}_{-3}$  &  305$^{+116}_{-65}$  & 0.28             &  10  \\    
                                    &  sech$^2$       &  10$^{+3}_{-2}$  &  412$^{+118}_{-66}$  & 0.63             &  10 \\  
\hline

$3.7 \leq P_{orb}$(h) $< 4.6$ (ALL) &  $\exp$         &  59$^{+8}_{-7}$  &  210$^{+25}_{-20}$  & 0.47             &  38  \\    
                                    &  sech$^2$       &  39$^{+5}_{-5}$  &  337$^{+32}_{-25}$  & 2.41             &  38 \\  
non-mCV                             &  $\exp$         &  35$^{+6}_{-5}$  &  253$^{+40}_{-31}$  & 1.03             &  25  \\    
                                    &  sech$^2$       &  24$^{+4}_{-4}$  &  390$^{+49}_{-36}$  & 2.46             &  25 \\  
mCV                                 &  $\exp$         &  23$^{+6}_{-5}$  &  170$^{+54}_{-33}$  & 0.01             &  13  \\    
                                    &  sech$^2$       &  16$^{+4}_{-3}$  &  250$^{+67}_{-39}$  & 0.01             &  13 \\  
\hline
$4.6 \leq P_{orb}$(h) $< 12$ (ALL) &  $\exp$          &  66$^{+9}_{-8}$  &  192$^{+20}_{-16}$  & 1.65            &  38  \\    
                                     &  sech$^2$      &  41$^{+6}_{-5}$  &  319$^{+26}_{-21}$  & 2.21            &  38 \\  
non-mCV                              &  $\exp$        &  48$^{+8}_{-6}$  &  213$^{+29}_{-23}$  & 0.81            &  30  \\    
                                     &  sech$^2$      &  31$^{+5}_{-4}$  &  342$^{+36}_{-28}$  & 1.72            &  30 \\  
mCV                                  &  $\exp$        &  11$^{+4}_{-3}$  &  271$^{+111}_{-60}$  & 0.17            &  8  \\    
                                     &  sech$^2$      &   8$^{+2}_{-2}$  &  412$^{+137}_{-71}$  & 0.61            &  8 \\  
\hline
\end{tabular}
}
\end{table}

Table 4 shows that the exponential function well represents the $z$-histogram 
constructed for all CVs (ALL) in the sample, dwarf novae (DN), non-magnetic 
CVs (non-mCV) and magnetic CVs (mCV), with scaleheights of 213, 191, 236 and 
173 pc, respectively. However, the $z$-histograms of nova-like stars (NL) and 
novae (N) are better fitted by the sech$^2$ function which gives scaleheights 
of 404 and 196 pc, respectively. It should be noted that the number of novae 
in the sample is too small to draw firm conclusions. Since the exponential function represents 
the vertical distribution of all CVs in the sample better, a comparison of 
$n_0$ with $N_{obs}$ for the model distribution produced using this function 
reveals that there are 30$\pm$8 CVs hiding in the Solar vicinity. Moreover, 
it is possible to estimate roughly the number of these missing systems in 
terms of the classes by comparing $N_{obs}$ with $n_0$ obtained from the 
exponential fits in Table 4: 17$\pm$6 dwarf novae, 8$\pm$4 nova-like stars 
and 4$^{+3}_{-2}$ novae. We performed Monte Carlo simulations to estimate 
the contribution of thick-disc CVs in the Solar neighbourhood to the Galactic 
model parameters since \cite{Aketal13} found that about 6 per cent of CVs in 
the Solar neighbourhood are members of the thick-disc population in the Galaxy. 
Assuming that 6, 8 and 10 per cent of CVs in the sample are the thick-disc CVs, 
our Monte Carlo simulations demonstrate that the effect of these systems to the 
scaleheights derived above is less than 4 per cent, which can be considered 
as a negligible contribution.

\begin{figure}
\begin{center}
\includegraphics[scale=0.22, angle=0]{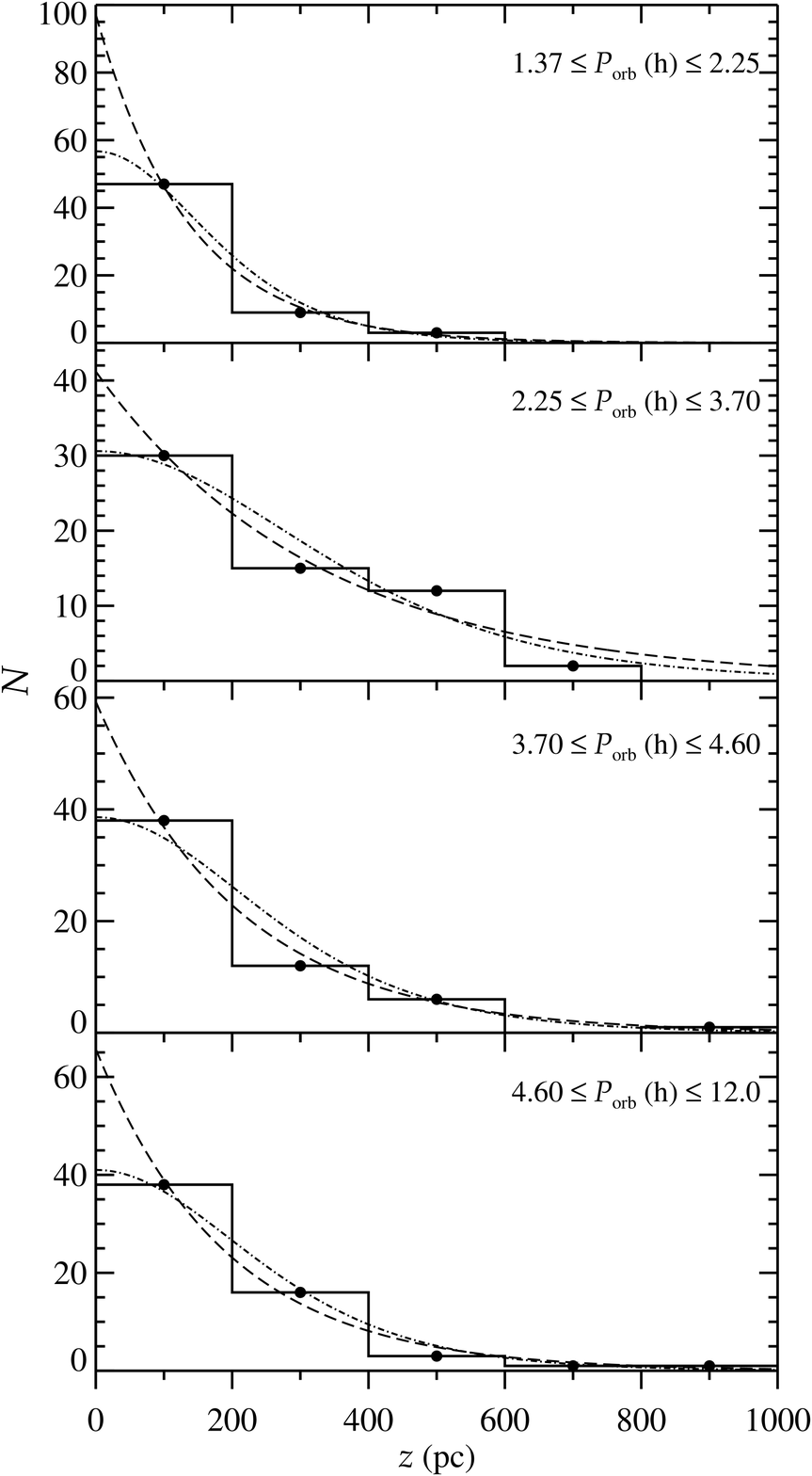}
\caption[] {\small The $z$-histograms for all CVs in the subgroups defined in 
terms of the orbital period. Lines are as described in Fig. 7.}
\end{center}
\end{figure}

The Galactic model parameters of CVs classified in terms of the orbital period 
were derived, as well. In order to define the limits of the orbital period 
intervals, we divided the CVs in the sample according to almost the same number 
of systems at four period intervals: 59 systems in the period interval 
$1.37 \leq P_{orb}$(h) $< 2.25$, 59 systems in $2.25 \leq P_{orb}$(h) $< 3.7$, 
57 systems in $3.7 \leq P_{orb}$(h) $< 4.6$ and 59 systems in 
$4.6 \leq P_{orb}$(h) $< 12$. By this classification, the total number of CVs to be 
used in deriving the Galactic model parameters drops to 234 as a limiting magnitude 
of 14 mag was set for CVs with $4.6 \leq P_{orb}$(h) $< 12$ and systems with 
$z \geq 1$ kpc were ignored to ensure the completeness of the subgroups, while 
the limiting magnitude of 16 selected above was maintained for CVs in the 
subgroups with $P_{orb} < 4.6$ h. Magnetic and non-magnetic systems in 
these period intervals were also considered as different subgroups. Note that 
the number of magnetic systems is very small for orbital periods above the 
orbital period gap. The $z$-histograms of CVs in these subgroups with their 
best fits are shown in Figs. 9-11. The Galactic model parameters are given 
in Table 5. The exponential function well represents the $z$-histograms constructed 
for the CV groups in Figs. 9-11, in general. Table 5 and Figs 9-11 show that the systems 
with shorter orbital periods ($2.25 \leq P_{orb}$(h) $< 3.7$) have larger scaleheights 
than CVs in the period range 
$3.7 \leq P_{orb}$(h) $< 12$. For the exponential function, the scaleheight 
increases from 192 to 326 pc while the orbital period decreases from 12 to 2.25 h. 
However, this trend is broken for the shortest orbital period CVs with 
$P_{orb} < 2.25$ h: the scaleheight suddenly decreases to 135 pc for systems with 
$P_{orb} < 2.25$ h. A similar trend is also found for the sech$^2$ function. 
The scaleheight for the sech$^2$ function changes from 319 to 444 pc for CVs 
with increasing periods in the range $2.25 \leq P_{orb}$(h) $< 12$, while it drops 
to 230 pc for CVs with 
$P_{orb} < 2.25$ h. The scaleheights estimated from the exponential and sech$^2$ 
functions for non-magnetic systems (non-mCV) show the same trend as found for 
all systems in the sample.

\begin{figure}
\begin{center}
\includegraphics[scale=0.22, angle=0]{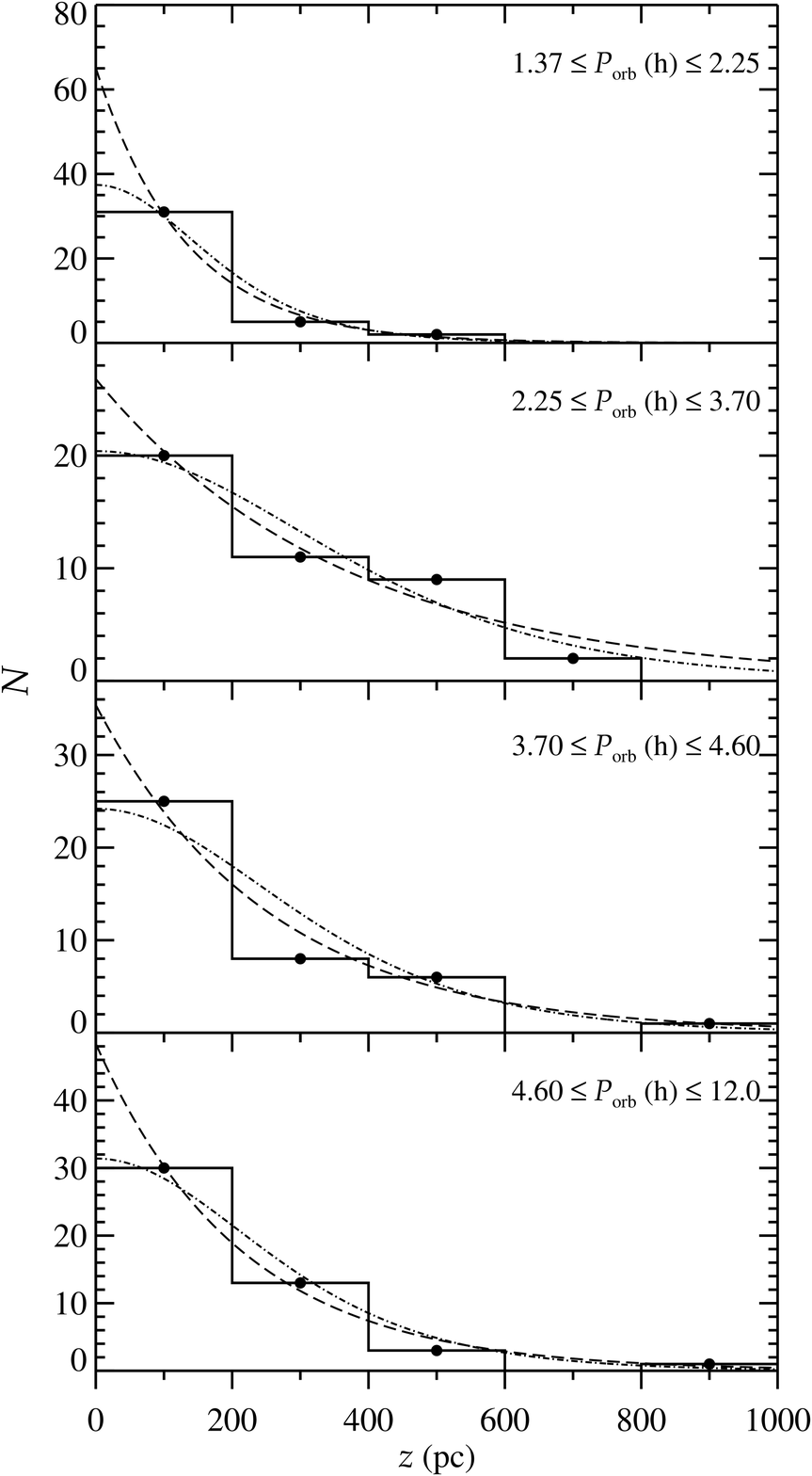}
\caption[] {\small The $z$-histograms for non-magnetic CVs in the subgroups 
defined in terms of the orbital period. Lines are as described in Fig. 7.}
\end{center}
\end{figure}

\begin{figure}
\begin{center}
\includegraphics[scale=0.22, angle=0]{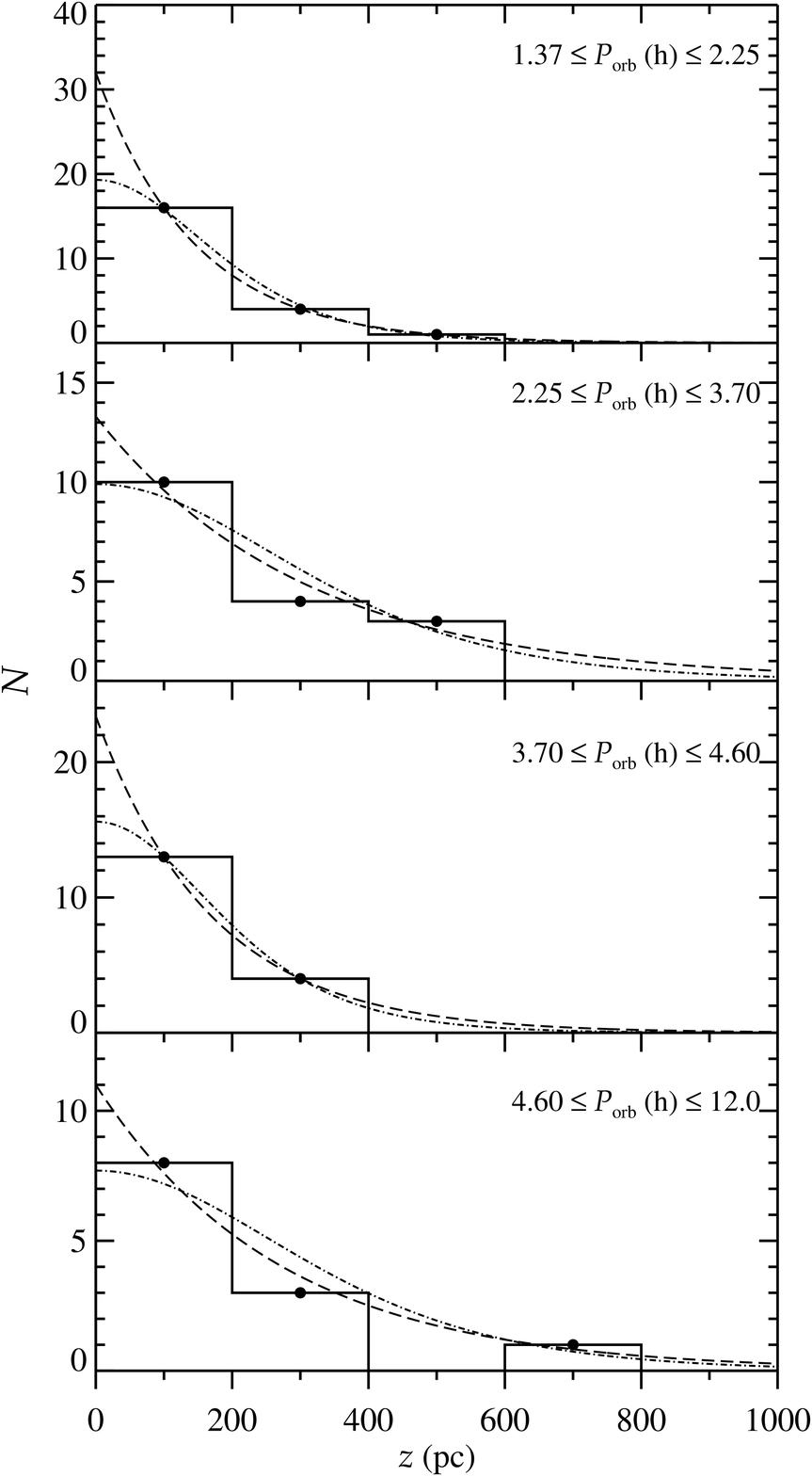}
\caption[] {\small The $z$-histograms for magnetic CVs in the subgroups 
defined in terms of the orbital period. Lines are as described in Fig. 7.}
\end{center}
\end{figure}

We derived the space density of CVs by dividing the number of stars in 
consecutive distances from the Sun to the corresponding partial spherical volumes: 
$D=N/\Delta V_{i,i+1}$  \citep{Biletal06a,Biletal06b} where $N$ is the number of 
stars in the partial spherical volume $\Delta V_{i,i+1}$ which is defined by 
consecutive distances $d_i$ and $d_{i+1}$ from the Sun. The logarithmic space 
density is expressed $D^*=\log D+10$. The logarithmic density functions of all 
CVs, non-magnetic CVs, magnetic CVs, dwarf novae, nova-like stars and novae in 
the Solar neighbourhood are shown in Fig. 12. In Fig. 12, $r^*$ denotes the 
centroid distance of the partial spherical volume: 
$r^*=[(d^{3}_{i}+d^{3}_{i+1})/2]^{1/3}$. The local space density is the space 
density estimated for $r^*=0$ pc. The local space densities obtained for all 
CVs and subgroups in the sample are listed in Table 6 which shows that the 
space density of CVs in the Solar neighbourhood is 5.58(1.35)$\times 10^{-6}$ pc$^{-3}$. 
Dwarf novae's (DN) space density is almost the same as found for all CVs in the sample: 
4.05(1.27)$\times 10^{-6}$ pc$^{-3}$. However, the space density of nova-like stars (NL) is 
only 34 per cent of the space density of DNs. The space density of magnetic 
(mCV) and non-magnetic CVs (non-mCV) are found to be very similar: 
3.13(0.77)$\times 10^{-6}$ and 2.45(0.70)$\times 10^{-6}$ pc$^{-3}$, respectively. 
The space densities of CVs in different period intervals are also given in 
Table 6. Table 6 demonstrates that the space density of CVs in the orbital period 
intervals above 2.25 h are almost the same with an average value of 
0.62$\times 10^{-6}$ pc$^{-3}$, while the space density of systems with 
$P_{orb} < 2.25$ h is about six times the space density of CVs with 
$P_{orb} > 2.25$ h.


\begin{figure}
\begin{center}
\includegraphics[scale=0.27, angle=0]{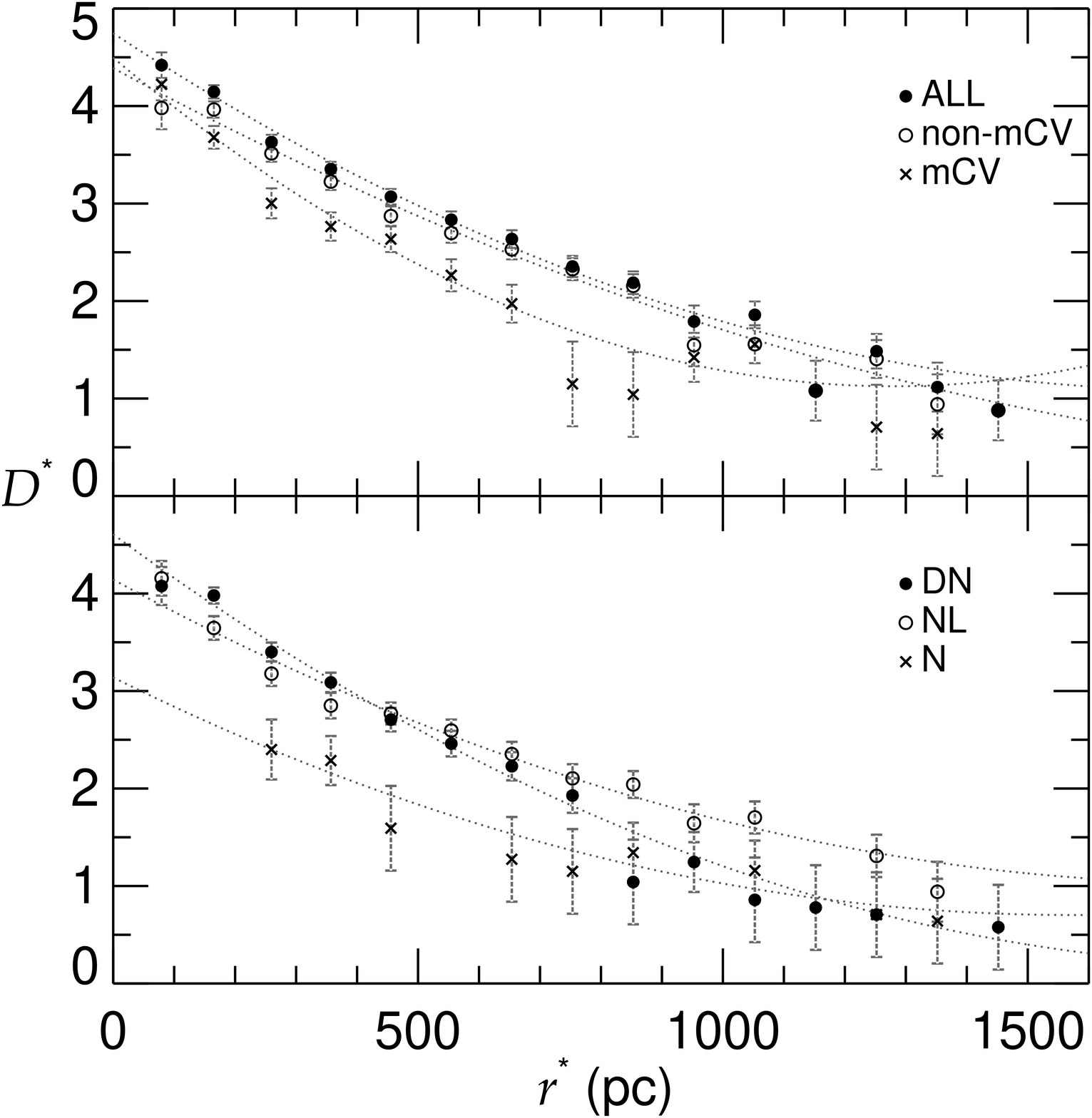}
\caption[] {\small The logarithmic density functions of all CVs (ALL), non-magnetic 
systems (non-mCV), magnetic systems (mCV), dwarf novae (DN), nova-like stars (NL) 
and novae (N) in the sample. Dotted lines represent polynomial fits applied to 
the data. }
\end{center}
\end{figure}


\begin{table}
\center
{\scriptsize
\caption{The local space densities of CVs. Symbols for subgroups are as in Table 3. 
The space densities of CVs classified in terms of the orbital period are also given.}
\begin{tabular}{lcclc}
\hline
Subgroup &  $D_0$($\times 10^{-6}$ pc$^{-3}$)     & ~~~~~~~~~~~~ &   Period Interval   & $D_0$($\times 10^{-6}$ pc$^{-3}$)  \\
\hline
ALL       &  5.58 $\pm$ 1.35     & ~~~~~~~~~~~~ &   $1.37 \leq P_{orb}$(h) $< 2.25$  & 3.73 $\pm$ 0.12  \\
DN        &  4.05 $\pm$ 1.27     & ~~~~~~~~~~~~ &   $2.25 \leq P_{orb}$(h) $< 3.7$  & 0.67 $\pm$ 0.03  \\
NL        &  1.39 $\pm$ 0.39     & ~~~~~~~~~~~~ &   $3.7 \leq P_{orb}$(h) $< 4.6$  & 0.64 $\pm$ 0.03  \\
N         &  0.14 $\pm$ 0.10     & ~~~~~~~~~~~~ &   $4.6 \leq P_{orb}$(h) $< 12$ & 0.54 $\pm$ 0.02  \\
non-mCV   &  2.45 $\pm$ 0.70     & ~~~~~~~~~~~~ &                                  &                  \\
    mCV   &  3.13 $\pm$ 0.77     & ~~~~~~~~~~~~ &                                  &                  \\
\hline
\end{tabular}
}
\end{table}

\subsection{Luminosity function}

We defined the luminosity function as the space density of objects in a certain 
absolute magnitude interval $M_{J1}-M_{J2}$ following \cite{Biletal06a,Biletal06b}. 
The partial spherical volume $\Delta V_{i,i+1}$ including the objects 
is defined by $d_i$ and $d_{i+1}$ distances which correspond to the bright and 
faint limiting apparent magnitudes of $J_0=10$ and $J_0=16$, respectively, for the 
absolute magnitude interval in question. The logarithmic luminosity functions $\phi$ 
are estimated for all systems, dwarf novae, nova-like stars, novae, non-magnetic 
systems and magnetic systems in the sample and listed in Table 7 and plotted in 
Fig. 13. An inspection of Table 7 shows that luminosity functions for all subgroups 
increase towards fainter absolute magnitudes. Table 7 indicates that dwarf 
novae and nova-like stars have similarly shaped luminosity functions. In addition, similar 
luminosity functions are also found for non-magnetic and magnetic systems. Novae 
have a luminosity function rather smaller than those estimated for all the subgroups.


\begin{figure}
\begin{center}
\includegraphics[scale=0.20, angle=0]{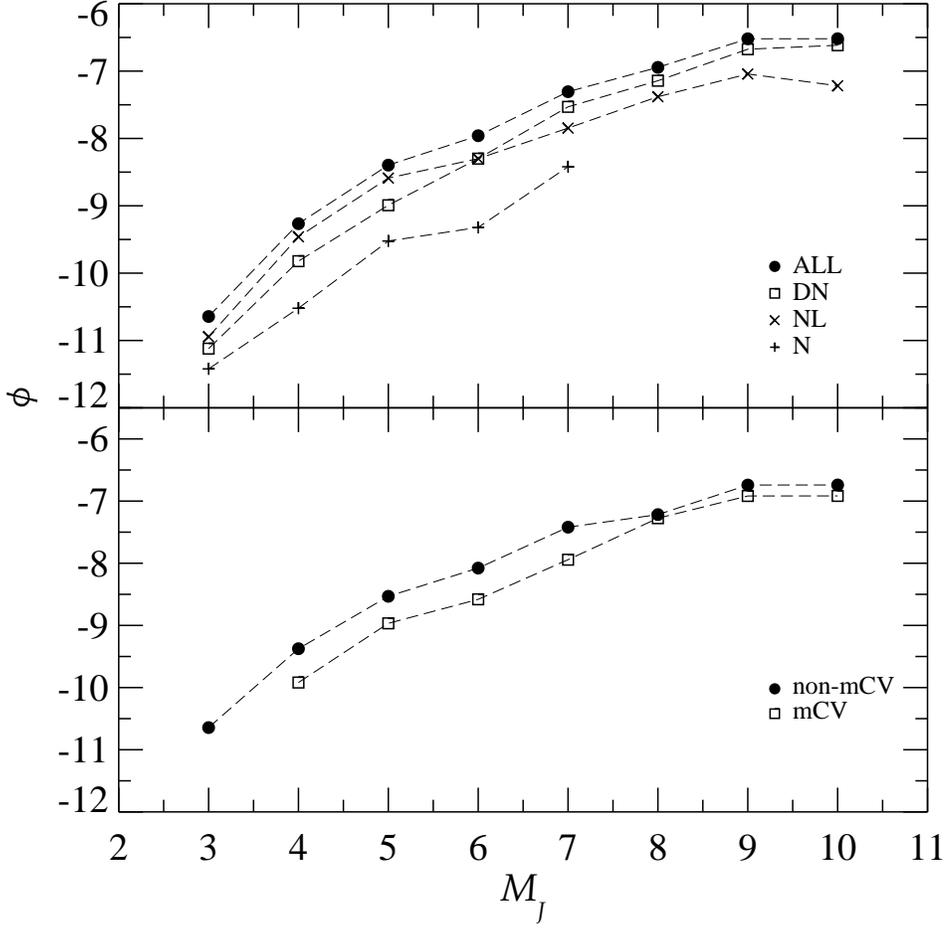}
\caption[] {\small The logarithmic luminosity functions of CVs in the sample. 
Denotes are as in Fig. 12.}
\end{center}
\end{figure}


\begin{table}
\setlength{\tabcolsep}{1pt}
\center
{\scriptsize
\caption{The logarithmic luminosity functions of CVs in the sample. $N$ is the 
number of stars in the $M_{J1}-M_{J2}$ absolute magnitude interval and $\phi$ the 
logarithmic luminosity function. Distance is in pc, volume pc$^{3}$.}
\begin{tabular}{ccccccccccccccc}
\hline
&&& \multicolumn{2}{c}{All Systems} 
& \multicolumn{2}{c}{Dwarf novae} 
& \multicolumn{2}{c}{Nova-like Stars} 
& \multicolumn{2}{c}{Novae}
& \multicolumn{2}{c}{non-Magnetic CVs} 
& \multicolumn{2}{c}{Magnetic CVs} \\
$M_{J1}-M_{J2}$	& $d_{1}-d_{2}$ & $\Delta V$ & N & $\phi$ & N & $\phi$ & N & $\phi$ & N & $\phi$ & N & $\phi$ & N & $\phi$  \\
\hline
(1.5, 2.5] & 398--6310 & 1.05(E12) &  1  & --- &  - & --- & 1 & ---  & - & --- & 1 & --- &  -  & ---          \\
(2.5, 3.5] & 251--3981 & 2.64(E11) & 6  & -10.64$\pm$1.90 &  2 & -11.12$\pm$3.40 & 3 & -10.94$\pm$2.70  & 1 & -11.42$\pm$5.00 & 6 & -10.64$\pm$1.90 & -  & ---             \\
(3.5, 4.5] & 158--2512 & 6.64(E10) & 36 & -9.27$\pm$0.70 & 10 & -9.82$\pm$1.40 & 23 & -9.46$\pm$0.90  & 2 &  -10.52$\pm$3.20 & 28 & -9.37$\pm$0.70 & 8 & -9.92$\pm$0.70\\
(4.5, 5.5] & 100--1585 & 1.67(E10) & 67 & -8.40$\pm$0.40 & 17 & -8.99$\pm$0.90 & 43 & -8.59$\pm$0.60  & 5 &  -9.52$\pm$1.80 & 49 & -8.53$\pm$0.40 & 18 & -8.97$\pm$0.40\\
(5.5, 6.5] & 63--1000  & 4.19(E9)  & 46 & -7.96$\pm$0.50 & 21 & -8.30$\pm$0.80 & 21 & -8.30$\pm$0.80  & 2 &  -9.32$\pm$2.90 & 35 & -8.08$\pm$0.50 & 11 & -8.58$\pm$0.50\\
(6.5, 7.5] & 40--631	   & 1.05(E9)  & 52 & -7.31$\pm$0.40 & 31 & -7.53$\pm$0.60 & 15 & -7.85$\pm$0.90  & 4 &  -8.42$\pm$1.80 & 40 & -7.42$\pm$0.40 & 12 & -7.94$\pm$0.40\\
(7.5, 8.5] & 25--398	   & 2.64(E8)  & 30 & -6.94$\pm$0.60 & 19 & -7.14$\pm$0.70 & 11 & -7.38$\pm$1.00  & -  & ---                & 16 & -7.22$\pm$0.60 & 14 & -7.28$\pm$0.60 \\
(8.5, 9.5] & 16--251	   & 6.62(E7)  & 20 & -6.52$\pm$0.60 & 14 & -6.67$\pm$0.80 & 6 & -7.04$\pm$1.20  & -  & ---                 & 12 & -6.74$\pm$0.60 & 8 & -6.92$\pm$0.60 \\
(9.5, 10.5]& 10--158   & 1.65(E7)  & 5  & -6.52$\pm$1.20 & 4  & -6.62$\pm$1.40 & 1 & -7.22$\pm$3.10  & -  & ---                  & 3 & -6.74$\pm$1.30 & 2 & -6.92$\pm$1.30 \\
\hline
\end{tabular}
}
\end{table}


\begin{figure}
\begin{center}
\includegraphics[scale=0.25, angle=0]{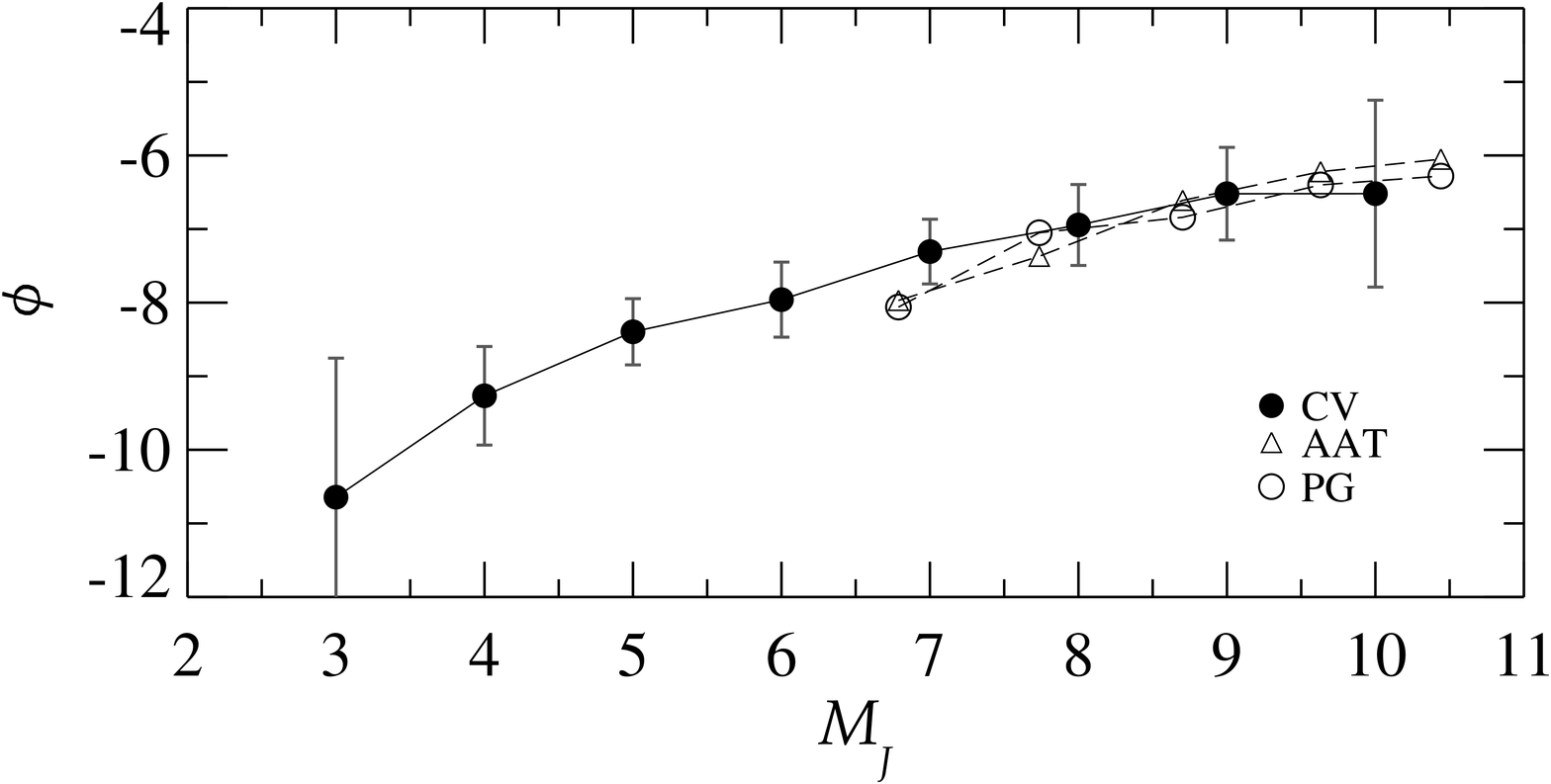}
\caption[] {\small Comparison of the luminosity functions of CVs obtained in 
this study with the luminosity functions of DA white dwarfs derived from the 
PG \citep{Flemingetal1986} and AAT \citep{Boyle1989} surveys . The best fits 
are obtained by dividing the luminosity functions of white dwarfs from 
PG and AAT surveys by a factor of 350-450 respectively.}
\end{center}
\end{figure}

For a comparison of the luminosity function of DA white dwarfs found from 
the Anglo Australian Telescope survey \citep[AAT,][]{Boyle1989} and Palomar 
Green survey \citep[PG,][]{Flemingetal1986} with the luminosity function of 
CVs derived in this study, the Johnson $M_{V}$ absolute magnitudes of CVs 
are transformed to 2MASS $M_{J}$ absolute magnitudes using Padova 
Isochrones \citep{Bresetal12}. Since the spatial distribution of CVs in Fig. 6 
shows that these systems belong to the old thin disc of the Galaxy, in general, 
an analytical relation between $M_{V}$ and $M_{J}$ by assuming a mass 
fraction of metals of $z=0.019$, a logarithmic surface gravity of $\log g>4$ 
and a mean age of $t=5$ Gyr \citep{Aketal10} are obtained in the selection 
of isochrones. The comparison of the luminosity functions of DA white dwarfs 
and CVs in the sample is shown in Fig. 14. The $\chi^{2}_{min}$ analysis 
show that the best fits between the luminosity functions of CVs and white 
dwarfs are obtained by dividing the luminosity functions of white dwarfs 
a factor of 450 and 350 for the PG and AAT surveys, respectively.

\section{Conclusions}

In order to derive the spatial distribution, Galactic model parameters and 
luminosity function of CVs, a new PLCs relation has been obtained using the 
2MASS and $WISE$ photometries. This new PLCs relation gives more reliable 
results than that in \cite{Aketal07}. A CV sample was collected using the 
new PLCs relation. The sample was limited in terms of the $J$-band magnitudes 
to ensure its completeness. The Galactic model parameters, space densities and 
luminosity functions derived in this study can be used to constrain the population 
models of CVs. Our conclusions can be summarized as follows:

\begin{itemize}
\item Estimated distances to the Sun are smaller than 1 kpc for CVs in the sample. 
CVs in the Solar neighbourhood are located in the thin-disc with $z$ distances lower 
than $\sim$700 pc, in general, a result consistent with \citep{Aketal13}. 

\item The exponential scaleheight of CVs in the sample is 213$^{+11}_{-10}$ pc. This value 
is in agreement with the exponential scaleheights of 190$\pm30$ and 160-230 pc suggested 
by \cite{Pat84} and \cite{vParetal96}, respectively. However, it is considerably larger 
than the exponential scaleheight of 158$\pm$14 pc derived by \cite{Aketal08}. 

\item If the $z$-distribution of all CVs in the sample is modeled by a sech$^2$ function 
\citep{Biletal06a,Biletal06b}, the scaleheight of CVs in the Solar 
neighbourhood is derived 326$^{+13}_{-12}$ pc. This scaleheight seems to be acceptable as 
\cite{Ganetal09} concluded that the scaleheight of CVs is very likely larger than the 190 pc 
found by \cite{Pat84}. Moreover, \cite{Pretal07b} argues that the 190 pc used by \cite{Pat84} can 
be suitable only for youngest CVs. This is a very plausible argument since the previous 
measurements for the scaleheight of CVs were based on the samples which are strongly biased 
towards bright objects and these measurements were made using only the exponential function. 
Thus, we suggest that the scaleheight of CVs in the Solar vicinity should be $\sim$330 pc and 
that the sech$^2$ function should be also considered in the CV population models. However, 
it must be keep in mind that the bias of the bright CVs in the sample can be dominant on 
the derivation of observational Galactic model parameters.

\item Our analysis show that the systems with short orbital periods (2.25-3.7h) have larger 
scaleheights than that estimated for CVs with $P_{orb} > 3.7$ h. Also, it seems that the 
observational scaleheights derived in this study are not in agreement with the scaleheights 
adopted by \cite{Pretal07b}. The trend of the scaleheight in terms of the orbital period 
encourages us to predict that the exponential and sech$^2$ scaleheights of the CVs 
with $P_{orb} < 2.25$ h are larger than 326 and 444 pc, respectively, even though the 
observed scale height of these (faint) systems is much smaller.

\item The logarithmic density functions of CVs show that the density functions derived in 
this study is in agreement with those shown in \cite{Aketal08}. We found that the space 
density of CVs in the Solar neighbourhood is 5.58(1.35)$\times 10^{-6}$ pc$^{-3}$, a result 
consistent with 
\citep{Wa74,Pat84,Pat98,Ri93,Schetal02a,ABetal05,Pretal07a,Aketal08,Revetal08,PrKn12,Pretal13}. 
However, the predicted values in the population synthesis studies are $10^{-5}-10^{-4}$ pc$^{-3}$  
\citep{RB86,dK92,Kol93,Pol96,Wiletal05,Wiletal07}. Although we used a fairly complete 
data sample, this comparison shows that there is still not a satisfactory agreement between the 
observational and predicted space densities of CVs.

\item Space densities of the CVs with $P_{orb} > 2.25$ h in the Solar neighbourhood are almost 
the same: 0.62$\times 10^{-6}$ pc$^{-3}$, in average. However, the space density of CVs with 
the orbital periods shorter than 2.25 h is about 6 times larger than that derived for 
longer period CVs. This result indicates that the faintest systems with distances greater 
than 200 pc could not be observed yet. 

\item Although it is not clear if masses of the white dwarfs in CVs are similar to those of 
isolated DA white dwarfs \citep{Sion1999}, our analysis based on the comparisons of 
the luminosity function of DA white dwarfs (Fig. 14) found from the AAT survey 
\citep{Boyle1989} and the PG survey \citep{Flemingetal1986} 
with the luminosity function of CVs show that the best fits are obtained by dividing the 
luminosity functions of white dwarfs by 450 and 350 for the PG and AAT surveys, respectively. 

\end{itemize}

\section{Acknowledgments}

This work has been supported in part by the Scientific and Technological Research 
Council of Turkey (T\"UB\.ITAK) 212T091. This work has been supported in part by Istanbul 
University: Project number 27839. We thank the anonymous referee for a through report and 
useful comments that helped improving an early version of the paper. This research has made 
use of the Wide-field Infrared Survey Explorer and NASA/IPAC Infrared Science Archive and 
Extragalactic Database (NED), which are operated by the Jet Propulsion Laboratory, California 
Institute of Technology, under contract 
with the National Aeronautics and Space Administration. This publication makes use of data products 
from the Two Micron All Sky Survey, which is a joint project of the University of Massachusetts and 
the Infrared Processing and Analysis Center/California Institute of Technology, funded by the 
National Aeronautics and Space Administration and the National Science Foundation. This research 
has made use of the SIMBAD, and NASA's Astrophysics Data System Bibliographic Services.

\end{document}